\PassOptionsToPackage{pdfpagelabels=false,final}{hyperref}
\documentclass[fleqn,usenatbib]{mnras}

\usepackage[T1]{fontenc}
\usepackage{ae,aecompl}

\usepackage{graphicx}	% Including figure files
\usepackage{amsmath}	% Advanced maths commands
\usepackage{amssymb}	% Extra maths symbols
\usepackage{xspace}		% Adds extra space if needed
\usepackage[modulo,switch]{lineno}
\usepackage[usenames,dvipsnames]{color}
\usepackage{bm}         % Extra bold maths symbols 

\newcommand{\BASTA}{{\tt BASTA}\xspace}

\newcommand{\Ktwo}{{\it K2}\xspace}
\newcommand{\Kepler}{{\it Kepler}\xspace}
\newcommand{\TESS}{{\sc TESS}\xspace}

\def\teff{T_{\rm eff}}
\def\feh{\rm[Fe/H]}
\def\meh{\rm[M/H]}
\def\logg{\log\,(g)}

\def\afe{\rm[\alpha/Fe]}
\def\msun{\rm M_{\odot}}
\def\dnu{\Delta\nu}
\def\num{\nu_\mathrm{max}}
\def\runo{r_{010}}

\usepackage{newtxtext,newtxmath}

\title[The BAyesian STellar Algorithm]{The BAyesian STellar Algorithm (\BASTA\thanks{The code is available via \url{https://github.com/BASTAcode/BASTA}}): a fitting tool for stellar studies, asteroseismology, exoplanets, and Galactic archaeology}

\author[V.~Aguirre B\o rsen-Koch et al.]{V. Aguirre B\o rsen-Koch$^1$\thanks{Formerly V. Silva Aguirre}, J.~L. R\o rsted$^1$\thanks{Formerly J.~R. Mosumgaard}, A.~B. Justesen$^1$, A. Stokholm$^{2,1}$, K. Verma$^1$,
\newauthor M.~L. Winther$^1$, E. Knudstrup$^1$, K.~B. Nielsen$^{1,3}$, C. Sahlholdt$^{4,1}$, J.~R. Larsen$^1$, 
\newauthor S. Cassisi$^{5,6}$, A.~M. Serenelli$^{7,8}$, L. Casagrande$^{9}$, J. Christensen-Dalsgaard$^1$, 
\newauthor G.~R. Davies$^{10,1}$, J.~W. Ferguson$^{11}$, M.~N. Lund$^1$, A. Weiss$^{12}$, and T.~R. White$^{13,1}$  
\\
$^1$Stellar Astrophysics Centre, Department of Physics and Astronomy, Aarhus University, Ny Munkegade 120, DK-8000 Aarhus C, Denmark\\
$^2$Dipartimento di Fisica e Astronomia, Universit\`a degli Studi di Bologna, Via Gobetti 93/2, I-40129 Bologna, Italy\\
$^3$Max Planck Institute for Astronomy, K\"{o}nigstuhl 17, D-69117 Heidelberg, Germany\\
$^4$Lund Observatory, Department of Astronomy and Theoretical Physics, Box 43, SE-221 00 Lund, Sweden\\
$^5$INAF-Astronomical Observatory of Abruzzo, Via M. Maggini sn, I-64100 Teramo, Italy\\
$^6$INFN - Sezione di Pisa, Largo Pontecorvo 3, 56127 Pisa, Italy\\
$^7$Instituto de Ciencias del Espacio (ICE, CSIC), Campus UAB, Carrer de Can Magrans, s/n, 08193 Cerdanyola del Valles, Spain\\
$^8$Institut d'Estudis Espacials de Catalunya (IEEC), Gran Capita 4, E-08034, Barcelona, Spain\\
$^9$Research School of Astronomy \& Astrophysics, Australian National University, ACT 2611, Australia\\
$^{10}$School of Physics and Astronomy, University of Birmingham, Birmingham, B15 2TT, UK\\
$^{11}$Physics Department, Wichita State University, Wichita, KS 67260-0032, USA\\
$^{12}$Max-Planck-Institut f\"{u}r Astrophysics, Karl Schwarzschild Strasse 1, 85748, Garching, Germany\\
$^{13}$Sydney Institute for Astronomy, School of Physics, University of Sydney, NSW 2006, Australia
}

\date{Accepted XXX. Received YYY; in original form ZZZ}

\pubyear{2021}

\begin{document}
\label{firstpage}
\pagerange{\pageref{firstpage}--\pageref{lastpage}}
\maketitle

\begin{abstract}
We introduce the public version of the BAyesian STellar Algorithm (\BASTA), an open-source code written in {\tt Python} to determine stellar properties based on a set of astrophysical observables. \BASTA has been specifically designed to robustly combine large datasets that include asteroseismology, spectroscopy, photometry, and astrometry. We describe the large number of asteroseismic observations that can be fit by the code and how these can be combined with atmospheric properties (as well as parallaxes and apparent magnitudes), making it the most complete analysis pipeline available for oscillating main-sequence, subgiant, and red giant stars. \BASTA relies on a set of pre-built stellar isochrones or a custom-designed library of stellar tracks which can be further refined using our interpolation method (both along and across stellar tracks/isochrones). We perform recovery tests with simulated data that reveal levels of accuracy at the few percent level for radii, masses, and ages when individual oscillation frequencies are considered, and show that asteroseismic ages with statistical uncertainties below 10\% are within reach if our stellar models are reliable representations of stars. \BASTA is extensively documented and includes a suite of examples to support easy adoption and further development by new users.
\end{abstract}

%%%%%%%%%%%%%%%%% BODY OF PAPER %%%%%%%%%%%%%%%%%%
\section{Introduction}
\label{sec:int}
%\linenumbers
Obtaining reliable properties of field stars is of paramount importance for many fields in astrophysics. %For instance, 
An accurate characterization of exoplanets requires precise knowledge of the parent star radius and mass, the ultimate fate and evolutionary remnant of a star can only be determined if we know its initial mass, and the study of the formation and evolution of our Galaxy is incomplete without the distribution of stellar ages across the Milky Way. These are just examples of how determining fundamental properties of stars has become the cornerstone of research for a variety of subjects.

Empirical methods to determine physical characteristics of stars are restricted to targets where e.g., years of monitoring are feasible (eclipsing binaries), or their bright apparent magnitude allows the measurement of the angular diameter (interferometry). This severely limits the number of stars where these techniques are applicable, and demands the development of methods in which some measured quantities (e.g., stellar effective temperature, surface composition, and luminosity) are compared to model predictions to infer stellar properties (such as age).

The advent of large-scale stellar surveys providing a myriad of data for thousands of stars across the Galaxy has led to the further development of algorithms that can combine different datasets to determine stellar properties. These algorithms vary in the approach used to extract the final parameters of stars (e.g., machine-learning, neural networks, Bayesian inference), the method to determine uncertainties (e.g., confidence intervals, Gaussian errors, Monte Carlo sampling), and the set of stellar tracks/isochrones considered in the analysis. Moreover, the possible combinations of input data are different for these codes: while some rely on spectroscopic, photometric, and astrometric data (e.g., {\tt StarHorse} \citep{2018MNRAS.476.2556Q}, {\tt MADE} \citep{2019MNRAS.484..294D}), others have the capability of including asteroseismic information (e.g., {\tt PARAM} \citep{2017MNRAS.467.1433R}, {\tt Isoclassify} \citep{2017ApJ...844..102H}, {\tt AIMS} \citep{2019MNRAS.484..771R}). The latter point is of key importance: due to their dependence on the internal stellar structure, reproducing the observed pulsation properties allows for a determination of the stellar radius, mass, and age of solar-type stars and red giants to a level of precision that cannot be achieved when fitting only atmospheric properties. Asteroseismology has therefore become an invaluable tool for a large variety of studies thanks to the rapidly increasing amount of data available since the launch of the space-missions CoRoT, {\it Kepler}, and TESS, which will continue to ramp up as future missions such as PLATO 2.0 \citep{2014ExA....38..249R} begin to acquire data.

The irruption of asteroseismology in the scene of stellar properties determination poses tremendous challenges to fitting algorithms due to the large variety of oscillation quantities that one can try to reproduce. In stars whose driving mechanism is stochastic excitation from their outer convective envelopes \citep[called solar-like oscillators, see e.g.][for a review]{2013ARA&A..51..353C}, the so-called global asteroseismic parameters can almost always be determined if oscillations are detected. If the data are of sufficient quality, individual frequencies of oscillation (or combinations of them) can be reproduced in main-sequence stars, as well as modes of mixed character that dominate the information content in subgiants. More evolved red giants present a much richer spectrum of pulsation, where state-of-the-art fitting algorithms reproduce only a subset of the observed frequencies while also considering a probe of the stellar core in the form of a gravity-mode period spacing. It is clear that fully exploiting the richness of asteroseismic data from solar-like oscillators requires algorithms capable of fitting a large variety of oscillation properties, which become relevant at different evolutionary stages and are strongly dependent on the data quality available. Moreover, these data should be supplemented with knowledge of the stellar effective temperature, a measurement of chemical composition, and ideally a determination of luminosity or absolute magnitude from astrometric and photometric data.

With this in mind we have developed the BAyesian STellar Algorithm (\BASTA), originally introduced in \citet{2015MNRAS.452.2127S}. \BASTA is a fitting tool written in {\tt Python} \citep{10.5555/1593511} designed to take advantage of the large variety of data obtained by large-scale ground-based surveys and space missions to precisely characterize stars. It has been built in a flexible way that allows the user to choose any combination of input data to be fit and, to the best of our knowledge, it is the fitting code that includes the largest number of global asteroseismic quantities and individual frequency diagnostics for solar-type stars, subgiants, and red giants. Other codes do not include information from individual oscillation frequencies (e.g., {\tt Isoclassify} and {\tt PARAM}), or glitch properties (e.g. {\tt AIMS}), or simply do not take any asteroseismic input (e.g., {\tt StarHorse} and {\tt MADE}). Moreover, \BASTA allows to simultaneously reproduce spectroscopic, photometric, astrometric, and asteroseismic data in a self-consistent manner, and is the only code where parallaxes can be fitted directly in addition to e.g., individual oscillation frequencies without the need of transforming the astrometric information into a luminosity estimate. \BASTA can run using publicly available compilations of stellar isochrones or tailor-made sets of evolutionary tracks with a wide combination of input physics.

\BASTA has been extensively used to determine stellar properties of both asteroseismic and non-seismic exoplanet host stars discovered by \Kepler \citep{2015MNRAS.452.2127S,2016NatCo...711201L,Bonomo_2019_Kepler107}, \Ktwo \citep{Johnson_2018_K2planets,Persson_2018_K2216,Van_Eylen_2018_HD89345,Hjorth_2019_K2290,Korth_2019_K2planets,Lund_2019_K293}, \TESS \citep{Gandolfi_2018_PiMen,Huber_2019_TESS_subgiant} and MASCARA \citep{Talens_2018_MASCARA2,Hjorth_2019_MASCARA3}. The sample of precise asteroseismic parameters originally derived in \citet{2015MNRAS.452.2127S} has enabled detailed studies of e.g. exoplanet eccentricities \citep{Van_Eylen_2015_ecc} and the radius gap \citep[][see also \citealt{Fulton_2017_radiusgap}]{Van_Eylen_2018_radius_gap}. Since then, the applications of \BASTA have been extended to a large variety of studies across fields of astrophysics such as characterisation of asteroseismic targets \citep[e.g.,][]{SilvaAguirre:2017eh,2020ApJ...889L..34A,2017ApJS..233...23S,2019MNRAS.489..928S}, Galactic archaeology \citep[e.g.,][]{2016MNRAS.455..987C,2018MNRAS.475.5487S,2019A&A...623A..60S,2020A&A...635A..58S,2020A&A...640A..81N}, open clusters \citep[e.g.,][]{2016MNRAS.463.2600L,2016ApJ...832..133S,2019A&A...622A.190A}, and the study of physical processes in stars such as rotation, convective overshoot, and magnetic activity \citep[e.g.,][]{2016Natur.529..181V,2017MNRAS.464.3713H,2017MNRAS.471.1012B}. It has also been shown to be one of the most accurate pipelines available in tests using artificial data of main-sequence stars \citep{2016A&A...592A..14R}.

In this paper we introduce the public version of \BASTA. We describe the Bayesian approach followed when determining stellar properties, the compilations of stellar track and isochrones available, the main features and capabilities included in the code, and present validation results of the fitting algorithm using artificial data.
\section{The Bayesian framework}\label{sec:bayes}
We use Bayesian statistics for stellar properties inference. In this framework, Bayes' theorem combines our prior knowledge about the model stellar parameters ${\bmath \Theta}$ (which includes e.g., mass, radius and age) with the information given by the data $\bmath D$ (such as measurements of effective temperature, metallicity and oscillation frequencies), to provide the posterior probability distribution of model stellar parameters,
\begin{equation}\label{eq:bayes}
P(\bmath{\Theta} | {\bmath D}) = \frac{P({\bmath D} | {\bmath \Theta}) P({\bmath \Theta})}{P({\bmath D})}.   
\end{equation}
Here, $P({\bmath D} | {\bmath \Theta})$ or the likelihood is the probability of observing the data given the model parameters, $P({\bmath \Theta})$ or the prior is the probability of parameters without seeing the data, and $P({\bmath D})$ or the evidence is the total probability of observing the data (which is a normalising constant).

We define the likelihood assuming Gaussian-distributed uncertainties on all observables except the magnitudes (see Section~\ref{sssec:phot}). \BASTA is developed with great emphasis on enabling the user to fit a variety of observables including the ones coming from spectroscopy (e.g. effective temperature, $\teff$, surface metallicity, $\feh$, and logarithm of surface gravity, $\logg$), astrometry and photometry (e.g. parallax, $\varpi$, and apparent magnitudes, $m_\zeta$), and asteroseismology (e.g. large frequency separation, $\dnu$, frequency of maximum power, $\num$, individual oscillation frequencies, $\nu$, and their combinations, $r_{01}$, $r_{10}$, $r_{02}$, $r_{012}$ and $r_{102}$). Note that there are correlations among some of these observables which we account for in the fitting process using the corresponding covariance matrix. To this purpose, we define the full likelihood as the product of likelihoods of groups of observables, $D_i$,
\begin{equation}\label{eq:likelihood}
P({\bmath D} | {\bmath \Theta}) = \prod_{i} P({\bmath D}_i | {\bmath \Theta}).
\end{equation}
The group likelihoods are computed using the expression (except for distances, see Section~\ref{sssec:phot} below),
\begin{equation}\label{eq:grouplike}
P({\bmath D}_i | {\bmath \Theta}) = \frac{1}{\sqrt{2\pi|\mathbfss{C}_i|}} \exp\left(-\chi_i^2 / 2\right),    
\end{equation}
where $|\mathbfss{C}_i|$ is the determinant of the covariance matrix, and 
\begin{equation}\label{eq:chi2}
\chi_i^2 = \frac{1}{N_i} \left(\bmath{O}_{i,{\rm obs}} - \bmath{O}_{i,{\rm mod}}\right)^{\rm T} \mathbfss{C}_i^{-1} \left(\bmath{O}_{{i,\rm obs}} - \bmath{O}_{i,{\rm mod}}\right).
\end{equation}
In Eq.~\ref{eq:chi2}, note the division by the number of observables, $N_i$. Although its inclusion is adhoc in a statistical sense, it can be useful in artificially reducing the weight of a group of observables (normally the individual oscillation frequencies). In \BASTA, the user can choose to turn off this division.

The user can specify an informative prior on stellar mass as given by the initial mass function (IMF). There are several versions of IMF included in \BASTA \citep{salpeter1955,millerscalo1979,kennicutt1994,scalo1998,kroupa2001,baldryglazebrook2003,chabrier2003}. To decrease the computation time, the user can pre-select a region of the grid for which the likelihoods are computed. This selection can be made on any available grid properties with user-defined tolerances. Technically, this is equivalent to assuming specific non-informative priors on certain stellar parameters.

We can use the computed posterior probability $P(\bmath{\Theta} | {\bmath D})$ to derive the marginalized posterior for any model stellar parameter $\theta$ using the expression,
\begin{equation}\label{eq:margpost}
P(\theta | {\bmath D}) = \int P(\theta, \bmath{\Theta}' | {\bmath D}) w_{\bmath {\Theta}} d{\bmath {\Theta}'},
\end{equation}
where ${\bmath {\Theta}}'$ represents all the model parameters except $\theta$. The weight $w_{\bmath {\Theta}}$ is used to account for the volume of the parameter space occupied by the model characterized by ${\bmath {\Theta}}$, i.e. half the distance to its neighbouring points in all dimensions used to generate the grid (see further details in section~\ref{sec:grids}).
\section{Grids of stellar models}\label{sec:grids}
The functionalities of \BASTA rely on the use of collections of stellar evolution tracks or isochrones to extract the properties of stars by means of Bayesian inference. In its current version \BASTA runs over publicly available sets of isochrones and tracks as well as custom-computed ones, which we process and store in Hierarchical Data Format version 5 (\textsc{HDF5}) and make available upon request. The functionalities and adopted input physics for each case are described in the following subsections.
\subsection{{\tt BaSTI} isochrones and tracks}\label{ssec:basti}
Stellar properties can be determined with \BASTA making use of the recently updated library "a Bag of Stellar Tracks and Isochrones" ({\tt BaSTI}\footnote{\url{http://basti-iac.oa-abruzzo.inaf.it}}). This compilation consists of evolutionary tracks and isochrones that are available for four distinct science cases defined by the inclusion of different physical processes as given in Table~\ref{tab:basti}. We give a brief description of its main features in this section, and refer the reader to \citet[][]{Hidalgo:2018dy} and \citet{2021ApJ...908..102P} for additional details.
\begin{table}
\centering
\begin{tabular}{c c c c}
\hline
Case & $\lambda_\mathrm{ov}$ & $D_\mathrm{diff}$ & $\eta$ \\%order following the path in the grid
\hline
1 & No & No & No \\
2 & Yes & No & No\\
3 & Yes & No & Yes\\
4 & Yes & Yes & Yes\\ 
\hline
\end{tabular}
\caption{Science cases of {\tt BaSTI} tracks and isochrones available in \BASTA. Columns show combinations of convective core overshooting ($\lambda_\mathrm{ov}$), microscopic diffusion ($D_\mathrm{diff}$), and mass loss ($\eta$). See text for details.}
\label{tab:basti}
\end{table}

The prescription for convective core overshooting consists of instantaneous mixing beyond the region formally defined by the Schwarzschild criterion, keeping the radiative temperature gradient in this region. In the case of main-sequence models with convective core, the overshoot region is defined by the distance $\lambda_\mathrm{ov}\times H_p$, where $H_p$ is the local pressure scale height and $\lambda_\mathrm{ov}$ is a free parameter. It has been set equal to 0.2, decreasing to zero when the mass decreases below a certain value. The approach used for decreasing $\lambda_\mathrm{ov}$ from its maximum value to zero depends on both the chemical composition and stellar mass \citep[see section~2.2 of][for details]{Hidalgo:2018dy}. During the core-helium burning stage, regardless of the considered science case, core mixing is modeled by accounting for semiconvection and suppression of breathing pulses.

The science case including atomic diffusion follow the prescription of \citet{1994ApJ...421..828T}, while mass-loss is taken into account in the formulation of \citet{1975MSRSL...8..369R} with an efficiency of the free parameter $\eta$ set to 0.3. The temperature stratification in the outer stellar layers is given by the \citet{1981ApJS...45..635V} formulation, but for the case of very-low mass stellar models for which outer boundary conditions based on accurate model atmospheres have been adopted \citep[see][for a detailed discussion on this topic]{Hidalgo:2018dy, 2021ApJ...908..102P}. 
For the adopted physical framework and solar heavy element distribution (see section~\ref{sssec:spectra} below), the calibration of a Standard Solar Model (SSM) sets the value of the mixing length parameter $\alpha_\mathrm{MLT}$, and of the initial solar metallicity and He abundance. At the solar age, the {\tt BaSTI} SSM\footnote{We note that the {\tt BaSTI} SSM has been calculated starting from the pre-main sequence and including diffusion of He and heavy elements} matches the solar luminosity and radius as well as the present $(Z/X)_\odot$ abundance ratio with a value for the mixing length parameter $\alpha_{\rm MLT}= 2.006$ \citep[we refer to][for details about the properties of the BaSTI SSM]{Hidalgo:2018dy}.

The {\tt BaSTI} stellar models are available both for the \citet{2011SoPh..268..255C} solar heavy element mixture, and for an $\alpha-$element enhanced mixture\footnote{The computation of a stellar model grid for an $\alpha-$depleted mixture is in progress.} ($\afe=+0.4$). For each selected metallicity, the corresponding initial helium abundance has been obtained by adopting an He-enrichment ratio equal to $\Delta Y/\Delta Z=1.31$, and a primordial He abundance equal to $Y_\mathrm{P}=0.247$. From the calibration of a SSM the resulting initial solar abundances are $Z_{\rm ini} = 0.01721$ and $Y_{\rm ini} = 0.2695$.
From the compilation of 21 metallicities initially available with the release of the {\tt BaSTI} models, we have increased the resolution in chemical composition and age by interpolating metallicity points for all science cases using the available online routine provided by the {\tt BaSTI} team.

The current release of the {\tt BaSTI} library also contains evolutionary tracks spanning masses from 0.1~$\msun$ to 15~$\msun$. For a subset of these (ranging from $\sim0.7~\msun$ to 4.0~$\msun$ with the lower limit being metallicity dependent) the full interior structure is also provided throughout the evolution. The number of structures stored\footnote{The complete database of interior structures is publicly available at the {\tt BaSTI} URL repository.} at each mass varies with metallicity, ranging from $\sim300$ individual models at the high-mass end to more than 9,000 structures at the low-mass end. These are stored in the standardised {\tt fgong} format described in the website\footnote{\url{https://github.com/vaguirrebkoch/aarhus_RG_challenge}} of the Aarhus Red Giants Challenge \citep{2020A&A...635A.164S}. The availability of interior structures allows us to compute oscillation frequencies and determine asteroseismic observables as described in Section~\ref{ssec:astero}.

Our grids of {\tt BaSTI} isochrones and tracks reach ages up to 16~Gyr, which allow us to properly construct the posterior distributions of old stars without risking the appearance of an edge effect \citep[see e.g.,][]{2014A&A...561A.125V,2015A&A...575A..12V}.
\subsection{Custom-computed evolutionary tracks}\label{ssec:tracks}
While the use of publicly available compilations of tracks and isochrones (such as {\tt BaSTI}) makes it easy to compare the \BASTA results with those from other fitting codes, it poses a limit on our flexibility to explore different combinations of input physics and numerical implementations. For this reason, we developed algorithms that calculate and process user-defined grids of evolutionary tracks with any combination of input physics and make them readily usable with \BASTA. The limitations on the input physics are given by the features available in an evolutionary code, and currently we support the usage of {\tt GARSTEC} \citep{2008Ap&SS.316...99W} and the latest publicly available version of MESA \citep{2011ApJS..192....3P,2013ApJS..208....4P,2015ApJS..220...15P}.

In the case of {\tt GARSTEC}, our running version of the code has experienced a number of developments since the published version of \citet{2008Ap&SS.316...99W}. Updated nuclear reactions from Solar Fusion II are available \citep{2011RvMP...83..195A}, and electron screening of nuclear reactions now also covers the intermediate regime using the prescriptions of \citet{1973ApJ...181..439D} and \citet{1973ApJ...181..457G}. The definition of convective boundaries follows the recipe of \citet{2014A&A...569A..63G}, and includes a treatment of semiconvective mixing as described in \citet{SilvaAguirre:2011jz}. {\tt GARSTEC} can couple on-the-fly distilled information from the {\tt Stagger} grid of 3D-hydrodynamical simulations of stellar atmospheres during the evolution  \citep{2018MNRAS.481L..35J,2018MNRAS.478.5650M,2019MNRAS.tmp.2585M,2019MNRAS.488.3463J}. The prescription in {\tt GARSTEC} for overshoot consists of a diffusive process with a diffusion constant given by
\begin{equation}\label{eqn:diff_ove} 
D(z)=D_0\exp{\left(\frac{-2z}{fH_P}\right)}\,,
\end{equation}
where the constant $D_0$ is derived from the MLT convective velocities, $z$ is the radial distance from the edge of the convective zone, $f$ is a free efficiency parameter, and $H_P$ is the pressure scale height. For small convective cores the overshooting efficiency is limited using a geometrical cut-off factor $g_\mathrm{cut}$ that scales the local pressure scale height as follows:
\begin{equation}\label{eqn:hp_ove} 
H_P=\min\left\{H_P,H_P \left(\frac{R_\mathrm{cz}f}{g_\mathrm{cut}H_P}\right)^2\right\}\,.
\end{equation}
Here $R_\mathrm{cz}$ is the radial thickness of the convective zone. The default value of the free parameter is $g_\mathrm{cut}=2$, and it can be modified to allow finer control over the size of small convective cores, a desired feature in e.g., studies constraining the extent of convective cores using asteroseismic data \citep{SilvaAguirre:2013in,Deheuvels:2016ek}.

We have two different approaches for grid sampling. In the first, which is the conventional approach, we compute tracks on a predefined mesh of stellar parameters. The mesh points are typically equally spaced along each parameter. The stellar model grids calculated in this manner are known as Cartesian grids. In the second approach, we sample the parameter space uniformly using a quasi-random number generator described in \citet{sobo67}. Note that quasi-random number generators perform better than pseudo-random number generators, specifically in high-dimensional spaces, as the latter provides more clumpy distributions. We refer to grids computed in this way as Sobol grids.

We can generate Cartesian and Sobol grids over a number of stellar parameters including mass ($M_{\rm ini}$), initial helium abundance ($Y_{\rm ini}$), initial metallicity ($[{\rm Fe}/{\rm H}]_{\rm ini}$), alpha enhancement ($\afe$), mixing-length ($\alpha_{\rm MLT}$), overshoot and mass-loss ($\eta$). The parameters used to generate the grid define the dimension of the weight $w_{\bmath {\Theta}}$ included in our marginalised posterior distribution (see Eq.~\ref{eq:margpost}). Note that we construct grids over $[{\rm Fe}/{\rm H}]_{\rm ini}$ (instead of initial metal mass fraction), because this quantity is well constrained by the observed metallicity, and hence allows convenient choice of the parameter space over which one needs to calculate a grid to model the observed star. We can either treat $Y_{\rm ini}$ as a free parameter similar to other stellar parameters, or determine it from the $[{\rm Fe}/{\rm H}]_{\rm ini}$ assuming values for the primordial helium abundance \citep[default $Y_p = 0.248$;][]{2020JCAP...03..010F}, and the helium-to-metal enrichment ratio (default $\Delta Y / \Delta Z = 1.4$). The default values can be changed by the user. In the case of {\tt GARSTEC}, we have implemented alpha enhanced and depleted mixtures of the \citet{2009ARA&A..47..481A} solar abundances, ranging from $\afe=-0.2$ to $\afe=0.6$ in steps of 0.1. We adopt consistent opacity tables from OPAL for high-temperatures \citep{1996ApJ...464..943I} and the \citet{2005ApJ...623..585F} opacities in the low-temperature regime.
\section{Available fitting parameters}\label{sec:input}
One of the core features of \BASTA is its ability to handle inhomogeneities in the available data as long as they are contained in the grid model parameters. \BASTA can easily deal with heterogeneous input and provide a robust set of stellar properties based on the likelihood of the models in its grids. In the following sections, we describe the quantities available for fitting asteroseismic, photometric, spectroscopic, and astrometric data. A complete and up-to-date list of code parameters can be found in the code documentation (available on {\tt GitHub}).
\subsection{Asteroseismology}\label{ssec:astero}
The wealth of data from main-sequence and red giant stars provided by asteroseismic space missions has driven the development of \BASTA towards the study of solar-like oscillators. These are stars whose oscillations are excited by the same mechanism as in the Sun, and comprise the vast majority of targets where asteroseismic quantities are available. For solar-like oscillators, there are several observables that the user can select to be reproduced by the models and fit with \BASTA. In all grids of stellar models currently supported (see Section~\ref{sec:grids}), theoretical oscillation frequencies have been computed using the Aarhus adiabatic oscillation package \citep[ADIPLS,][]{2008Ap&SS.316..113C}.  
\subsubsection{Global asteroseismic quantities}\label{sssec:scal}
The number of asteroseismic properties that can be extracted from the power spectrum of a given star depends on the length of the observations, the target's apparent magnitude, and its evolutionary state (since the time-scale for oscillations scales with the intrinsic luminosity, see e.g., \citet{Kjeldsen:1995tr,Kjeldsen:2011de}). If the data reveal the signal of the oscillations, two basic seismic observables that can be readily extracted are the average large frequency separation $\langle\dnu\rangle$ and the frequency of maximum power $\num$. These quantities, also known as the global asteroseismic parameters, scale approximately with stellar properties as follows:
\begin{equation}\label{eqn:sca_dnu} 
\left(\frac{\langle\dnu\rangle}{\langle\dnu_\odot\rangle}\right)^{2} \simeq \frac{\bar{\rho}}{\bar{\rho}_\odot}\,,
\end{equation}
\begin{equation}\label{eqn:sca_num}
\frac{\num}{\nu_{\mathrm{max},\odot}} \simeq \frac{M}{\msun}{\left(\frac{R}{{\rm R_\odot}}\right)^{-2}} \left(\frac{\teff}{T_{\mathrm{eff},\odot}}\right)^{-1/2}\,,
\end{equation}
where $\langle\dnu_\odot\rangle$, $\nu_{\mathrm{max},\odot}$, and $T_{\mathrm{eff},\odot}$ are the values measured in the Sun. As $\langle\dnu\rangle$ and $\num$ are normalised to these reference solar values, they can be specified by the user. By default, \BASTA adopts $\langle\dnu_\odot\rangle=135.1\,\mu$Hz and $\nu_{\mathrm{max},\odot}=3090\,\mu$Hz from \citet{Huber:2011be}.

There is extensive literature devoted to the testing and validating the scaling relations by using independent constraints in stellar masses and radii from e.g., binary stars, clusters, interferometry, and parallaxes \citep[see e.g., ][to name a few]{SilvaAguirre:2012du,Huber:2012iv,Miglio:2012dm,White:2013bu,2016ApJ...832..121G,2017ApJ...844..102H,2018MNRAS.476.1931S,2018MNRAS.481L.125S,2018MNRAS.476.3729B}. There is general agreement that the scaling relations are accurate to within a few percent, but as it is clear from Eqs.~\ref{eqn:sca_dnu} and~\ref{eqn:sca_num} these extrapolations from the solar values do not take into account variations with chemical composition nor the evolutionary stage of the star.

To address some of these issues, several prescriptions have been proposed to correct the $\langle\dnu\rangle$ scaling relation, while the equation for $\num$ is purely empirical and no corrections across the range of interest in $\teff$ and $\feh$ have been derived yet \citep[but see][for an initial explanation on the theory behind Eq.~\ref{eqn:sca_num}]{Belkacem:2011hm}. We have implemented in \BASTA four additional determinations of the average large frequency separation that aim at further decreasing the level of systematic deviation in the scaling relation Eq.~\ref{eqn:sca_dnu}. A detailed comparison between the performance of each prescription can be found in \citet{2019ApJ...879...33V}.

The first two determinations consist of a linear fit as a function of radial order to the model individual frequencies of $\ell=0$ weighted by a Gaussian centred at $\num$, the only difference being the adopted Full Width at Half Maximum of $0.25\num$ \citep{White:2011fw} or $0.66\num^{0.88}$ \citep{2012A&A...537A..30M}. The average large frequency separation $\langle\dnu\rangle$ is determined from the slope of this fit and is meant to mimic as close as possible the manner in which this quantity is derived from the observations. This average large frequency separation is available for the grids of stellar models where we compute the individual oscillation frequencies (see Sections~\ref{ssec:basti} and~\ref{ssec:tracks}). As a by-product of this calculation we obtain the dimensionless offset $\epsilon$ that can be used to correct for differences between the observed and model radial order (see section~\ref{sssec:mmodes} below).

The other two determinations are those of \citet{2016ApJ...822...15S} and \citet{2017ApJS..233...23S} who, based on the same principle of the \citet{White:2011fw} approach, have computed a correction factor across the Hertzprung-Russell Diagram  for the value of $\langle\dnu\rangle$ obtained from Eq.~\ref{eqn:sca_dnu} that depends on the mass, metallicity, effective temperature, and evolutionary state (hydrogen-core or-shell burning or core-helium burning) of the star. These corrections can be computed for any grid of stellar tracks or isochrones, as they are independent of the availability of theoretical oscillation frequencies.

As a final remark, we note that when defining the solar reference values $\langle\dnu_\odot\rangle$ and $\nu_{\mathrm{max},\odot}$ in Eqs.~\ref{eqn:sca_dnu} and~\ref{eqn:sca_num} there is an implicit assumption that fitting a target with those values of $\langle\dnu\rangle$ and $\num$ (and solar temperature and metallicity) will result in a star of 1~M$_\odot$ and 1~R$_\odot$ (but not necessarily solar age, as this depends on the input physics used to construct the models). To ensure this level of consistency, all theoretical values of $\langle\dnu\rangle$ in our grids of models are re-scaled by a factor given by the fraction $\langle\dnu_\odot\rangle$/$\langle\dnu_\odot\rangle_\mathrm{grid}$, where $\langle\dnu_\odot\rangle_\mathrm{grid}$ is the average large frequency separation computed from the individual $\ell=0$ modes of a solar model computed with the same input physics of the corresponding grid using the FWHM of \citet{White:2011fw}.
\subsubsection{Individual oscillation frequencies}\label{sssec:mmodes}
If the time-series of observations is of sufficient signal-to-noise ratio and resolution, it is possible to extract individual oscillation frequencies characterized by radial order $n$ and angular degree $\ell$ \citep[see e.g.,][]{2016MNRAS.456.2183D,2017ApJ...835..172L}. For stellar disk-integrated observations, as it is the case of the space missions, geometrical cancellation suppresses the signal from all modes except those with low degree $\ell \le 3$. If the asymptotic theory can be applied to describe the oscillation \citep[e.g.,][]{Tassoul:1980dl}, modes of odd and even degree are separated by $\sim\langle\Delta\nu\rangle/2$. As a result, the observed oscillation spectrum contains one mode of each degree $\ell=1,2,3$ between two consecutive $\ell=0$ modes. Departures from this asymptotic description occur e.g., in the presence of mixed modes (see Section~\ref{sssec:smixed}).

Most current stellar evolutionary models use a rudimentary description of the outer-most layers of stars with convective envelopes \citep[e.g., the mixing-length theory of][]{BohmVitense:1958vy}, leading to systematic frequency shifts in the oscillation modes when compared to observations \citep{ChristensenDalsgaard:1996gf}. In order to correct for this effect, a so-called surface correction needs to be applied to the model frequencies. There are three prescriptions for the surface corrections implemented in \BASTA and all include power-law dependence on frequency. They aim at obtaining the corrected model frequencies $\nu_{n,\ell}^{\textup{cor}}$ from the original model frequencies $\nu_{n,\ell}^{\textup{model}}$ by determining the corresponding fitting coefficients of the power law correction to make them as close as possible to the observed frequencies $\nu_{n,\ell}^{\textup{obs}}$.

The first prescription is the empirical power-law correction from \cite{Kjeldsen:2008kw}:
\begin{equation}
r\nu_{n,\ell}^{\textup{cor}} - \nu_{n,\ell}^{\textup{model}} = \frac{a}{Q} \left(\frac{\nu_{n,\ell}^{\textup{model}}}{\nu_0}\right)^b,
\label{eqn:k08}
\end{equation}
in which $a$ and $r$ are the fitting coefficients, $b$ is a fixed exponent, and $\nu_0$ is a reference frequency, typically chosen to be the frequency of maximum power $\nu_\mathrm{max}$. $Q$ is the ratio between the inertia of the mode and the theoretical inertia that a radial ($\ell=0$) mode would have at that frequency, determined by linear interpolation.

The other implemented corrections are the two physically motivated surface corrections from \cite{Ball:2014gx}, first their \emph{cubic} correction:
\begin{equation}
\nu_{n,\ell}^{\textup{cor}} - \nu_{n,\ell}^{\textup{model}} = \frac{a_3 (\nu_{n,\ell}^{\textup{model}} / \nu_0)^3 }{I},
\label{eqn:bg14cubic}
\end{equation}
and secondly their \emph{combined} correction, adding the cubic correction from above to an inverse term:
\begin{equation}
\nu_{n,\ell}^{\textup{cor}} - \nu_{n,\ell}^{\textup{model}} = \frac{a_{-1} (\nu_{n,\ell}^{\textup{model}} / \nu_0)^{-1} + a_3 (\nu_{n,\ell}^{\textup{model}} / \nu_0)^3 }{I}.
\label{eqn:bg14}
\end{equation}
In these corrections, $a_{-1}$ and $a_{3}$ denotes the fitting coefficients, and $I$ is the scaled mode inertia, typically normalized at the stellar surface as \citep{Aerts:2010uw}, 
\begin{equation}
I = \frac{4 \pi \int_0^R \left[| \tilde{\xi_r} |^2 + \ell(\ell+1) | \tilde{\xi_h}|^2\right] \rho_0 r^2\,\textup{d}r}{M\left[| \tilde{\xi_r(R)} |^2 + \ell(\ell+1) | \tilde{\xi_h(R)}|^2\right]},
\label{eq:modeinertia}
\end{equation}
where $\xi_r$ and $\xi_h$ are the radial and horizontal components of the displacement, $\rho_0$ is the unperturbed stellar density, $M$ is the total stellar mass, and $R$ is the photospheric radius.

\begin{figure*}
\includegraphics[width=\textwidth]{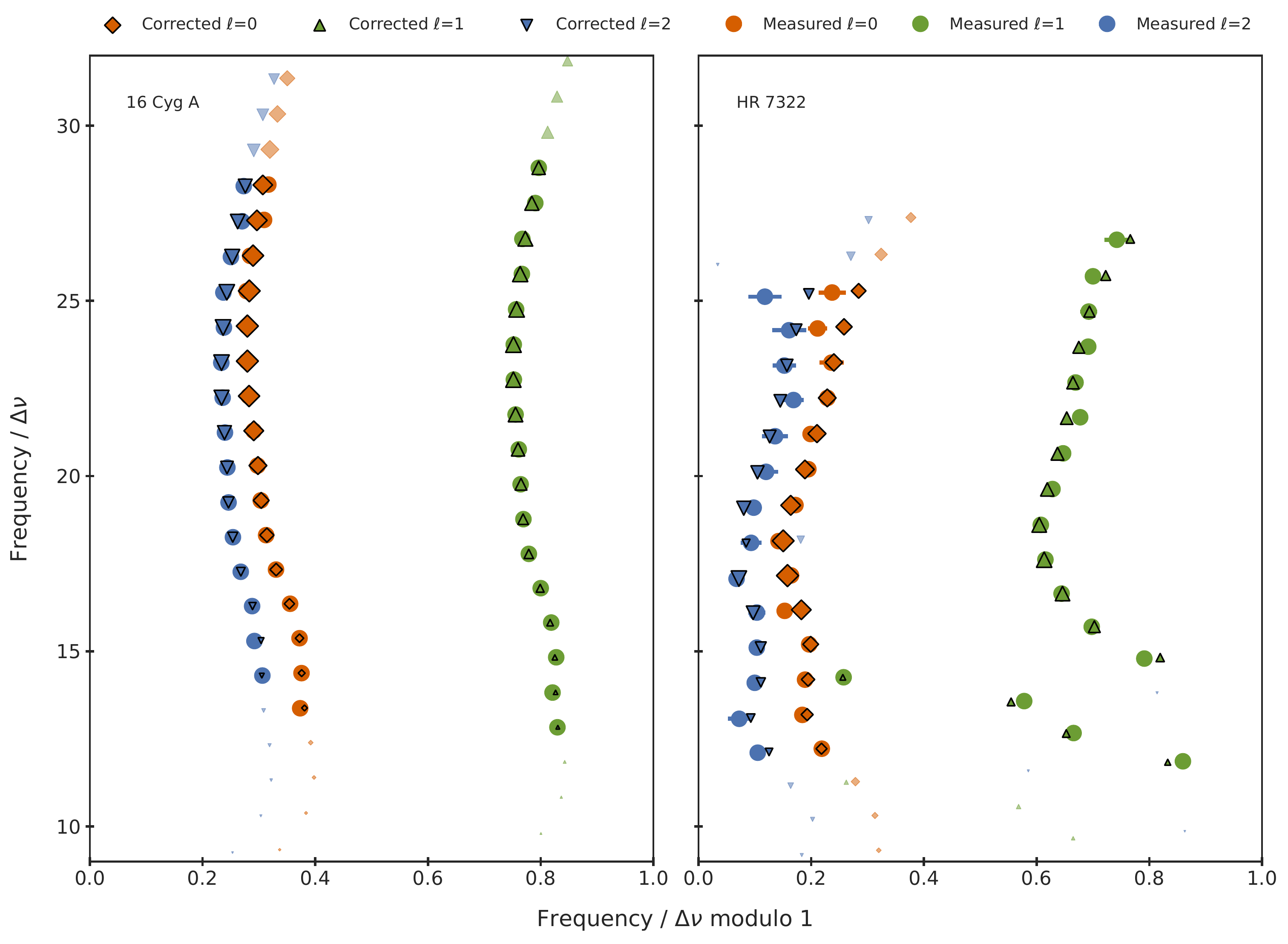}
\caption{\'Echelle diagram of the main-sequence star 16~Cyg~A and the subgiant star HR~7322. The coloured circles show the observed oscillation modes (red: $\ell=0$, green $\ell=1$, blue: $\ell=2$) from \citet{2017ApJ...835..172L} and \citet{2019MNRAS.489..928S}, respectively,  while the coloured symbols with a black outline show the modes predicted from model with the highest likelihood (same colour coding), corrected using the combined surface correction from \citet{Ball:2014gx} and scaled using the observed $\Delta \nu$ of the star. The size of the symbols from the model are scaled inversely with their normalized mode inertias: the larger the symbol, the greater the probability of the mode being observed. The lighter coloured symbols with no outline are not matched to any observation, but still predicted by the model.}
\label{fig:frequencyfitting}
\end{figure*}

When fitting individual oscillation frequencies, it is necessary to correctly match each observed mode with its corresponding theoretical counterpart of identical radial order $n$ as the surface correction and thus the computation of the likelihood of the model depend on the difference in frequency between the observed and modelled frequency of the same radial order and angular degree. Identifying the radial order in the observations of main-sequence stars is relatively straight forward using the dimensionless offset $\epsilon$ \citep{2012ApJ...751L..36W}, and \BASTA has an option for the user to apply a suitable correction to the radial order of the observed frequencies, if desired. An example of the best fit model found by \BASTA when fitting the individual oscillation frequencies of the \Kepler target 16~Cyg~A is shown in Fig.~\ref{fig:frequencyfitting}.
\subsubsection{Frequency combinations}\label{sssec:ratios}
Combinations of frequencies have been long used in asteroseismic analysis to isolate the signature of a given stellar region and extract detailed information about the structure of a star \citep[e.g.,][]{Roxburgh:2003bb,OtiFloranes:2005ii,Cunha:2007gs,SilvaAguirre:2011jz}. A simple example of this is the large separation between modes of same angular degree and consecutive overtone, $\Delta\nu_{\ell}(n)=\nu_{n,\ell}-\nu_{n-1,\ell}$, that is related to the mean stellar density (cf., Eq.~\ref{eqn:sca_dnu}). In \BASTA we have included a number of combinations as fitting parameters such as the small frequency separation:
\begin{equation}\label{eqn:d02}
d_{02}(n)=\nu_{n,0}-\nu_{n-1,2}\,,
\end{equation}
the 5-point small frequency separations:
\begin{equation}\label{eqn:d01}
d_{01}(n)=\frac{1}{8}(\nu_{n-1,0}-4\nu_{n-1,1}+6\nu_{n,0}-4\nu_{n,1}+\nu_{n+1,0})
\end{equation}
\begin{equation}\label{eqn:d10}
d_{10}(n)=-\frac{1}{8}(\nu_{n-1,1}-4\nu_{n,0}+6\nu_{n,1}-4\nu_{n+1,0}+\nu_{n+1,1})\,,
\end{equation}
the frequency separation ratios:
\begin{equation}\label{eqn:r02}
r_{02}(n)=\frac{d_{02}(n)}{\Delta\nu_{1}(n)}
\end{equation}
\begin{equation}\label{eqn:r01}
r_{01}(n)=\frac{d_{01}(n)}{\Delta\nu_{1}(n)}\,,
\end{equation}
\begin{equation}\label{eqn:r10}
r_{10}(n)=\frac{d_{10}(n)}{\Delta\nu_{0}(n+1)}\,,
\end{equation}
and the set of ratios $\runo$, $r_{012}$, and $r_{102}$:
\begin{eqnarray}\label{eqn:rsets}
r_{010} &=& \{r_{01}(n),r_{10}(n),r_{01}(n+1),r_{10}(n+1),...\}\,,\\
r_{012} &=& \{r_{01}(n),r_{02}(n),r_{01}(n+1),r_{02}(n+1),...\}\,,\\
r_{102} &=& \{r_{02}(n),r_{10}(n),r_{02}(n+1),r_{10}(n+1),...\}\,.
\end{eqnarray}

There are strong correlations between combinations including five individual frequencies, and \citet{Deheuvels:2016ek} showed that the set $\runo$ results in almost singular covariance matrices with large condition numbers that can lead to overfitting the data as recently suggested by \citet{2018arXiv180807556R}. Instead, the latter study recommends the usage of $r_{01}$ or $r_{10}$ in combination with $r_{02}$ to form the series $r_{012}$ or $r_{102}$, respectively. Here, the underlying problem is in the use of conventional numerical methods to estimate the inverse of covariance matrices, which turns out to be highly inaccurate if these are ill-conditioned. We overcome this issue by using the Moore-Penrose pseudo-inverse \citep[see e.g.][]{strang2006linear} of the covariance matrices in the likelihood. The pseudo-inverse is calculated using singular value decomposition and sets singular values to zero when these are below a threshold that is defined relative to the largest singular value.

\begin{figure*}
\includegraphics[width=\textwidth]{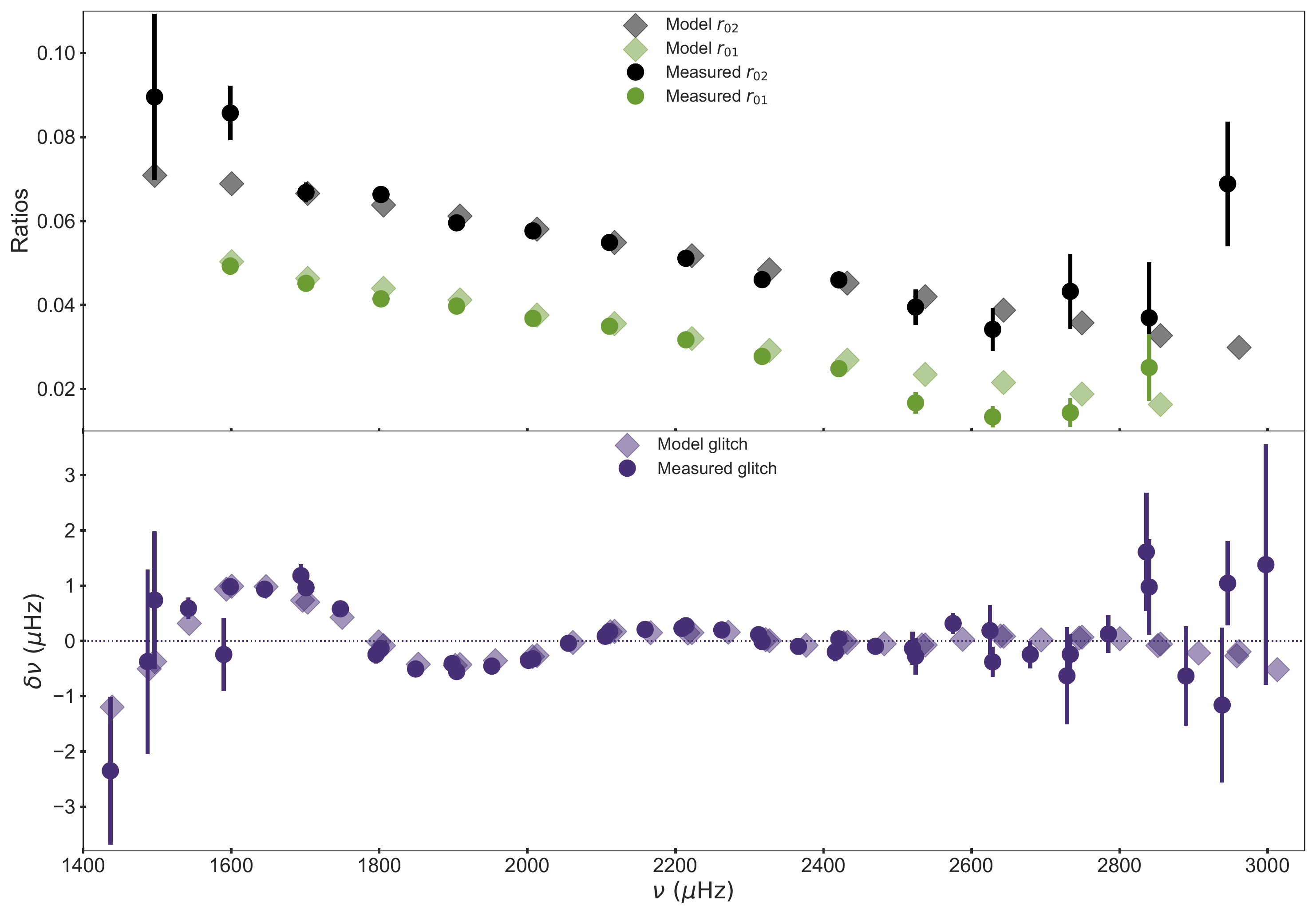}
\caption{Ratios and glitch signatures for 16~Cyg~A as a function of oscillation frequencies. In the top panel, the black and green circles represent the observed ratios $r_{02}$ and $r_{01}$, respectively, while the black and green diamonds show the corresponding quantities for the best-fitting model (see the legend). In the bottom panel, the circles represent the sum of the observed glitch signatures from the helium ionization zone and the base of convection zone, while the diamonds show the same for the best-fitting model.}
\label{fig:ratioGlitch}
\end{figure*}

The approach devised in \BASTA to handle frequency combinations attempts to give the user as much flexibility as possible while keeping the statistical approach robust, and allows fitting any of these quantities ($d_{01}$, $d_{10}$, $d_{02}$, $r_{01}$, $r_{10}$, $r_{02}$, $r_{010}$, $r_{012}$, $r_{102}$) as desired. In all cases the user only needs to provide the individual oscillation frequencies and \BASTA will calculate the needed combinations. It is possible to supply the code with the necessary correlations across terms, and if these are not given then \BASTA calculates them by doing 10,000 Monte Carlo realisations drawn from random Gaussian distributions of the individual frequencies. An example of the fit to the frequency ratios obtained for 16~Cyg~A is shown in the top panel of Fig.~\ref{fig:ratioGlitch}.
\subsubsection{Acoustic glitches}\label{sssec:glitches}
There are regions inside solar-like oscillating stars where the sound speed changes at length scales substantially shorter than the local wavelengths of the acoustic waves. These regions are known as acoustic glitches in the stellar structure, and the two most prominent are the helium ionization zone and the base of the envelope convection zone. Glitches in the acoustic structure of stars leave tiny signatures in the observed oscillation frequencies currently detectable from space borne missions \citep[see e.g.][]{migl10,mazu12} as well as from ground-based facilities \citep[see e.g.][]{bedd10,grun17}. The bottom panel of Fig.\@~\ref{fig:ratioGlitch} shows an example of glitch signatures as seen in the oscillation frequencies of 16 Cyg A observed by the \Kepler satellite.

The detection of the glitch signatures can provide useful information about stellar interior, e.g. they can be used to measure the location of the base of envelope convection zone as well as to infer the surface helium abundance \citep[see e.g.][]{mazu14,verm14a,verm17,verm19a,verm19b}. The perturbation to the oscillation frequencies due to acoustic glitches can be derived using the asymptotic theory of stellar oscillations \citep[see e.g.][]{houd07},
\begin{eqnarray}
\delta\nu &=& A_{\rm He} \nu e^{-8\pi^2\Delta_{\rm He}^2\nu^2} \sin(4\pi\tau_{\rm He}\nu + \psi_{\rm He})\nonumber\\
&+&\frac{A_{\rm CZ}}{\nu^2} \sin(4\pi\tau_{\rm CZ}\nu + \psi_{\rm CZ}),
\label{eqn:glh}
\end{eqnarray}
where the two terms on the right-hand side represent contributions from the helium and base of the convection zone glitches, respectively. The parameters $A_{\rm He}$, $\Delta_{\rm He}$, $\tau_{\rm He}$ and $\psi_{\rm He}$ depend on the properties of the helium ionizing layers, whereas $A_{\rm CZ}$, $\tau_{\rm CZ}$ and $\psi_{\rm CZ}$ depend on the properties of the base of the convection zone. The parameter $\tau_{\rm CZ}$ is of particular importance as it provides an estimate of the acoustic depth of the base of the convective envelope. Another quantity of interest is the average amplitude of helium glitch signature,
\begin{eqnarray}
\langle A_\nu \rangle &=& \frac{\int_{\nu_1}^{\nu_2} A_{\rm He} \nu e^{-8\pi^2\Delta_{\rm He}^2\nu^2} d\nu}{\int_{\nu_1}^{\nu_2} d\nu}\nonumber\\
&=& \frac{A_{\rm He} [e^{-8\pi^2\Delta_{\rm He}^2\nu_1^2} - e^{-8\pi^2\Delta_{\rm He}^2\nu_2^2}]}{16\pi^2\Delta_{\rm He}^2[\nu_2 - \nu_1]},
\label{amplitude}
\end{eqnarray}
which has been used in the past to measure the envelope helium abundance of solar-type stars.

We have implemented in \BASTA the capability to fit the observed glitch parameters. The helium and convection zone glitch parameters are extracted from oscillation frequencies using the Method A of \citet{verm17,verm19a}. Briefly, in Method A, we modelled the total oscillation frequency,
\begin{equation}
f(n, \ell) = \sum_{k=0}^{4} b_k(\ell) n^k + \delta\nu,
\label{eqn:total_freq}
\end{equation}
where the first term represents the smooth component of oscillation frequencies while the second term arises from glitches (see Eq.~\ref{eqn:glh}). The polynomial coefficients $b_k(\ell)$ are determined together with the glitch parameters by fitting Eq.~\ref{eqn:total_freq} to the oscillation frequencies using nonlinear optimization method based on Broyden-Fletcher-Goldfarb-Shanno (BFGS) algorithm \citep[see e.g.][]{flet87}. To estimate the uncertainties in the glitch parameters, we repeat the fitting process using 1,000 realisations of the individual oscillation frequencies and estimate the full covariance matrix. The uncertainties in the glitch parameters correspond to the square root of the diagonal terms in the matrix, which are consistent with the error bars obtained from the Hessian matrix.

In the current implementation, we do not use the parameters associated with the base of convection zone glitch in the stellar properties determination as it is typically difficult to reliably extract them from the contemporary frequency precision for two reasons: (1) the small amplitude of the base of convection zone signature, and (2) the issue caused by aliasing \citep{mazu01}. Having said that, it is straightforward to modify the current implementation to use the glitch parameters related to the base of convection zone.
\subsubsection{Mixed modes and the frequency matching routine}\label{sssec:smixed}
As stars evolve beyond the core-hydrogen burning phase their oscillation pattern develop irregularities: when the helium core contracts as a product of stellar evolution, the frequency of the g-modes increases and interactions between the p-mode behaviour near the surface and the g-mode behaviour near the core take place. Modes can experience mixed properties and exchange nature, causing these so-called mixed modes to deviate from the regular oscillation pattern of the p-modes and thus to be visible in an \'echelle diagram as avoided crossings.

Mixed modes have a substantial diagnostic potential as they are sensitive to properties of the stellar core. However, mixed modes complicate the analysis of individual frequencies as this departure from simple asymptotic theory results in the presence of more than one non-radial mode of a given angular degree between two consecutive $\ell=0$ modes, and some of these theoretically predicted mixed modes do not reach observable amplitudes. This complicates the matching of model modes to observed modes which can lead to incorrect computations of the surface correction and of the likelihood evaluation of the given model.

To address this issue we note that the amplitude of a mode can be roughly estimated from the mode inertia (see Eq.~\ref{eq:modeinertia}). The frequency fitting procedure in \BASTA uses a mode matching algorithm, where the observed modes are matched to their most-likely counterpart in the model using their frequencies as well as their inertias as a proxy of the likelihood of observability. We describe the matching procedure and assumptions in the following paragraphs.

The frequency matching routine is based on mode counting. Even though the extracted frequencies can be uncertain at times as they depend on the power spectrum background and systematic effects in the pipeline, these effects do not change the relative ordering of the modes. An $\ell=1$ mode and an $\ell=0$ mode will not exchange order due to effects such as the surface effect. Due to the physical nature of the radial modes, avoided crossings do not occur in the pattern of the $\ell=0$ modes. We therefore use the observed radial modes to define a number of frequency intervals. If higher angular degree modes are observed outside the frequency range encompassed by the radial modes, the frequency binning is extended to lower and higher frequencies in steps of $\Delta \nu$.

\BASTA counts how many modes of a given angular degree $\ell\ne0$ are present in each frequency bin between two consecutive radial modes. When there is an equal number of model and observed modes, the modes are matched based only on frequency. We note in passing that this is the typical case for main-sequence stars that do not present avoided crossings. If there are more observed modes than modelled modes in a given range, the model is rejected as it fails to accurately describe the observed pattern.

If, between two $\ell = 0$ modes, $j$ modelled modes exists with inertias $I_1 < \dots < I_j$ and $k$ observed modes of same angular degree with $j>k$, we need to select $k$ modelled modes to match one-to-one to the observed modes. This is usually the case when avoided crossing takes place in the model, but not all mixed modes have high enough amplitudes to be observed. Intuitively, one might think to just pick the $k$ modelled modes with the lowest inertia as they should have the highest likelihood of being observed. However, small differences in inertia might cause this to result in a miss-matching. Instead, \BASTA selects two inertia thresholds $a$ and $b$ (with $a<b$) and subdivides the modelled modes into a set $A$ with inertias less than $a$, a set $B$ with inertia between $a$ and $b$, and a set $C$ for inertias greater than $b$. The set $A$ thus contains modes that are likely to be detected, while set $C$ contains modes that are unlikely to be observed, and set $B$ contains modes that are somewhere in between.

By picking all modes in $A$ and a subset of $B$ such that $k$ modes are selected in total, the modes can be matched one-to-one in frequency to the observed modes. Specifically, the thresholds $a$ and $b$ are chosen based on the $k$'th smallest modelled inertia $I_k$: $a$ is $I_k / 10$ and $b$ is $10 I_k$. This ensures that $|A| \le k \le |A| + |B|$, where $|A|$ is the number of modes in the set. Each possible match is evaluated based on the total $\mathrm{L}_1$ distance in frequency space and the subset of $B$ that minimises this metric is chosen. This match between the observed and modelled modes will then be used in the following surface effect and model likelihood computation.

Figure\@~\ref{fig:frequencyfitting} shows the \'echelle diagrams for two examples of asteroseismic fitting to individual frequencies using this matching algorithm. The code can nicely reproduce the observed oscillation pattern of the main-sequence star 16~Cyg~A, and also follow the rapid evolution of a distinct dipole mixed mode in the bright F6 subgiant star HR~7322. Further examples of matching the mixed-mode pattern in subgiant stars are given in Section~\ref{sec:tests} and Appendix~\ref{app:validation} below.
\subsubsection{Period spacing}\label{sssec:DP}
The observed power spectra of red-giant stars exhibit a complex pattern due to the presence of mixed modes. As described above, mixed modes behave as gravity modes in the inner regions of the star and as acoustic modes in the outer layers, and hence their observations provide a unique opportunity to probe the conditions deep in the stellar core \citep[see e.g.][]{2011Sci...332..205B,2011Natur.471..608B,2012A&A...540A.143M}. The detection of the mixed modes makes it possible to measure the gravity mode period spacing, and currently measurements of the asymptotic period spacing for the dipole modes are available for several thousands of {\it Kepler} red-giant stars \citep[see e.g.][]{2013ApJ...765L..41S,2014A&A...572L...5M,2015MNRAS.447.1935D,2016A&A...588A..87V}.

We can use the observed dipole mode asymptotic period spacing in \BASTA as a quantity to be fitted. The corresponding asymptotic period spacing for the models is computed according to the formula,
\begin{equation}
\Delta P_1 = \sqrt{2} \pi^2 \left(\int \frac{N}{r} dr \right)^{-1},
\label{eqn:dp1}
\end{equation}
where $N$ is the Brunt-V\"{a}is\"{a}l\"{a} frequency, $r$ the radial coordinate and the integration is performed over the radiative interior. A few necessary considerations regarding the integral in Eq.~\ref{eqn:dp1}: (1) the integrand has a numerical singularity, and (2) tabulated values of $r$ have variable step size. These make rectangle and trapezoidal rules for the integration inaccurate (particularly if the step size is not very small), and the Newton-Cotes formulas with higher order accuracy inapplicable. For this reason, we use an adaptive Gauss-Kronrod quadrature method (7-points Gauss rule combined with 15-points Kronrod rule) to compute the integral with high accuracy \citep{kronrod1965nodes}. This requires the values of the integrand at intermediate $r$ (other than the tabulated values), which is obtained using the basis spline interpolation.
\subsection{Atmospheric properties}\label{ssec:classical}
\subsubsection{Surface chemical composition}\label{sssec:spectra}
Fitting the observed surface chemical composition requires certain assumptions about the relation between measured number density ratios of a given element and the metal mass fraction used to construct stellar models. For a given grid of evolutionary tracks or isochrones characterised by hydrogen, helium, and metal mass fractions ($X+Y+Z=1$), \BASTA assumes the following mapping:
\begin{align}
\meh & = \log_{10}\left(\frac{(Z/X)_\mathrm{model}}{(Z/X)_\odot}\right) \label{eq:meh}\\
%\meh & = \feh+\log_{10}\left(C\times10^{\afe}+(1-C)\right)\label{eq:alphascale} \\  
\feh & = \meh-corr\left(\afe\right)\,.\label{eq:alphascale}
\end{align}
The equations above depend on the solar heavy element distribution adopted to construct the grid, and a correction factor corr$\left(\afe\right)$ which is determined by comparing stellar tracks with and without the inclusion of alpha-elements enhancement \citep[see][]{1993ApJ...414..580S}. These correction factors, valid when all alpha-elements are equally enhanced, change according to the considered solar mixture. We have determined them for the fixed alpha enhancements of $\afe=+0.4$ in the \citet{1993oee..conf...15G} and \citet{1998SSRv...85..161G} solar mixtures, as well as all the available values of $\afe$ for the \citet{2009ARA&A..47..481A} solar composition (see Section~\ref{ssec:tracks}). We adopt corr$\left(\afe\right)=0.3016$ as given by \citet{2021ApJ...908..102P} when dealing with the \citet{2011SoPh..268..255C} compilation, which is appropriate for an alpha-enhancement value of $\afe=+0.4$.

Under these assumptions, all custom-constructed grids used in \BASTA are mapped from the input $\feh_\mathrm{ini}$ and $\afe$ into $\meh$ using Eq.~\ref{eq:alphascale}, and then the ratio $(Z/X)_\mathrm{model}$ is determined using Eq.~\ref{eq:meh}. The {\tt BaSTI} models on the other hand follow the inverse procedure where we use their initial mass fractions of $X$ and $Z$ as given in their database to determine $\meh$ following Eq.~\ref{eq:meh}, and $\feh$ is then calculated for the corresponding case of $\afe$ using Eq.~\ref{eq:alphascale} and the solar mixture of \citet{2011SoPh..268..255C}. This procedure ensures that all surface abundance ratios are computed in a consistent manner, but it is up to the user to ensure that the combination of input values of $\feh_\mathrm{ini}$ and $\afe$ are reasonable for observed stars. We note in passing that the surface abundance by mass of elements such as ${}^3{\rm He}$, ${}^{12}{\rm C}$, ${}^{13}{\rm C}$, ${}^{14}{\rm N}$, and ${}^{16}{\rm O}$ are also available in our custom-computed grids.
\subsubsection{Synthetic photometry, parallaxes, and distances}\label{sssec:phot}
All grids of stellar models available in \BASTA (cf., section~\ref{sec:grids}) can be mapped from the theoretical H-R diagram to various photometric systems using bolometric corrections (BCs) tables provided by \citet{Hidalgo:2018dy}. This allows us to determine synthetic magnitudes in more than 15 photometric systems including those that are of relevance for asteroseismic and exoplanet studies (e.g., {\it Kepler} and {\it TESS} passbands), compilations of bright stars (Tycho-2 and Hipparcos), large photometric surveys (e.g., 2MASS, Skymapper, sloan, and VISTA), and naturally all Gaia data releases. A list of the photometric systems currently available in \BASTA is provided in Table~\ref{tab:photcolors}.

\BASTA supports the inclusion of parallaxes as a fitting parameter together with at least one apparent magnitude $m_\zeta$. 
The evaluation is then done by comparing the grid-model absolute magnitude $M_\zeta$, computed from the bolometric luminosity and effective temperature using bolometric corrections, to the measured absolute magnitude computed from the apparent magnitude $m_\zeta$, an estimate of extinction, and the distance modulus from the observed parallax $\mu=5\log(d)-5$. The computation of the observed absolute magnitude undergoes multiple transformations and thus the assumption of this value being normally distributed like the other fitting parameters is weak. Instead this parameter is evaluated by constructing the likelihood distribution of $M_\zeta$ and including it in the calculation of the posterior using Eq.~\ref{eq:likelihood}.

The procedure when parallax is included as a fitting parameter is as follows. In the first step, \BASTA constructs prior distributions in distance and apparent magnitude using their observed values. The distance distribution is analytically calculated from the measured parallax and its associated uncertainty using the exponentially decreasing space density prior with a scale length of $1.35$~kpc described in \citet{2015PASP..127..994B} and \citet{Astraatmadja:2016kd}. For the apparent magnitude $m_\zeta$, we assume a normal distribution with the observed values of $m_\zeta$ and its uncertainty as the mean and standard deviation, respectively.

In the second step, \BASTA samples over the distance and apparent magnitude distributions to calculate the reddening. When multiple distributions are considered it is difficult to properly sample the tails of all distributions if one just draws samples following the distributions and let the number density of the samples determine the probability density function. If we draw samples from two normal distributions, then the odds of drawing values at e.g. 3 standard deviations away from the mean in both distributions is less than 1 in 100,000. This means that the tails of the resulting distribution would be artificially steep. To overcome this, \BASTA draws the samples systematically over a large range of values for each given parameter.

For a number $N$ of samples in distance, \BASTA draws half of them linearly across a range of $d\in [10^{(-0.4)},10^{(4.4)}]$~pc, and the other half as quantiles of the normal distribution around the mode of the distance distribution derived from the observed parallax as described above. Similarly, we produce $K$ apparent magnitude samples distributing half of them linearly across the limiting magnitudes published by each relevant survey (or else assume $m_\zeta\in[-10,25]$), and the rest from a normal distribution around the observed apparent magnitude.

For each of the $N\times K$ pairs of distance and apparent magnitude we determine the reddening E($B-V$) using the latest version of the {\tt Bayestar} dust map \citep[currently {\tt Bayestar19}, see][]{Green_2019}, and if the target falls outside of its coverage we apply the simpler map of \citet{1998ApJ...500..525S}. Since {\tt Bayestar} provides multiple estimates of the colour excess at each distance, we determine the individual reddening values using the median and standard deviation of those samples. To transform the reddening values into absorption $A_\zeta$ estimates at a given filter $\zeta$, we use extinction coefficients from Table~6 in \citet{Schlafly:2011iu} assuming the \citet{1989ApJ...345..245C} relation $A_\zeta=3.1\times$E($B-V$). For filters not contained in that compilation, we default to the temperature and metallicity dependent extinction coefficients from \citet{Casagrande:2014ez}.

We compute the absolute magnitudes $M_\zeta$ from our $N\times K$ groups of distance, apparent magnitude, and absorption using the distance modulus. This sample is converted into a probability distribution function by weighting each obtained absolute magnitude by the underlying observed distance and apparent magnitude probability distributions. This distribution is then included in the computation of the posterior (see Eq.~\ref{eq:likelihood}). We note that in cases where the parallax uncertainty is smaller than 5\% and the absolute magnitude prior is symmetric (i.e., if the distance between the median and each quantile is within a predefined threshold), we fit a Gaussian function to the probability distribution of absolute magnitudes and use the analytical expression in the computation of the likelihood.

\begin{figure*}
%    \centering
    \includegraphics[width=84mm]{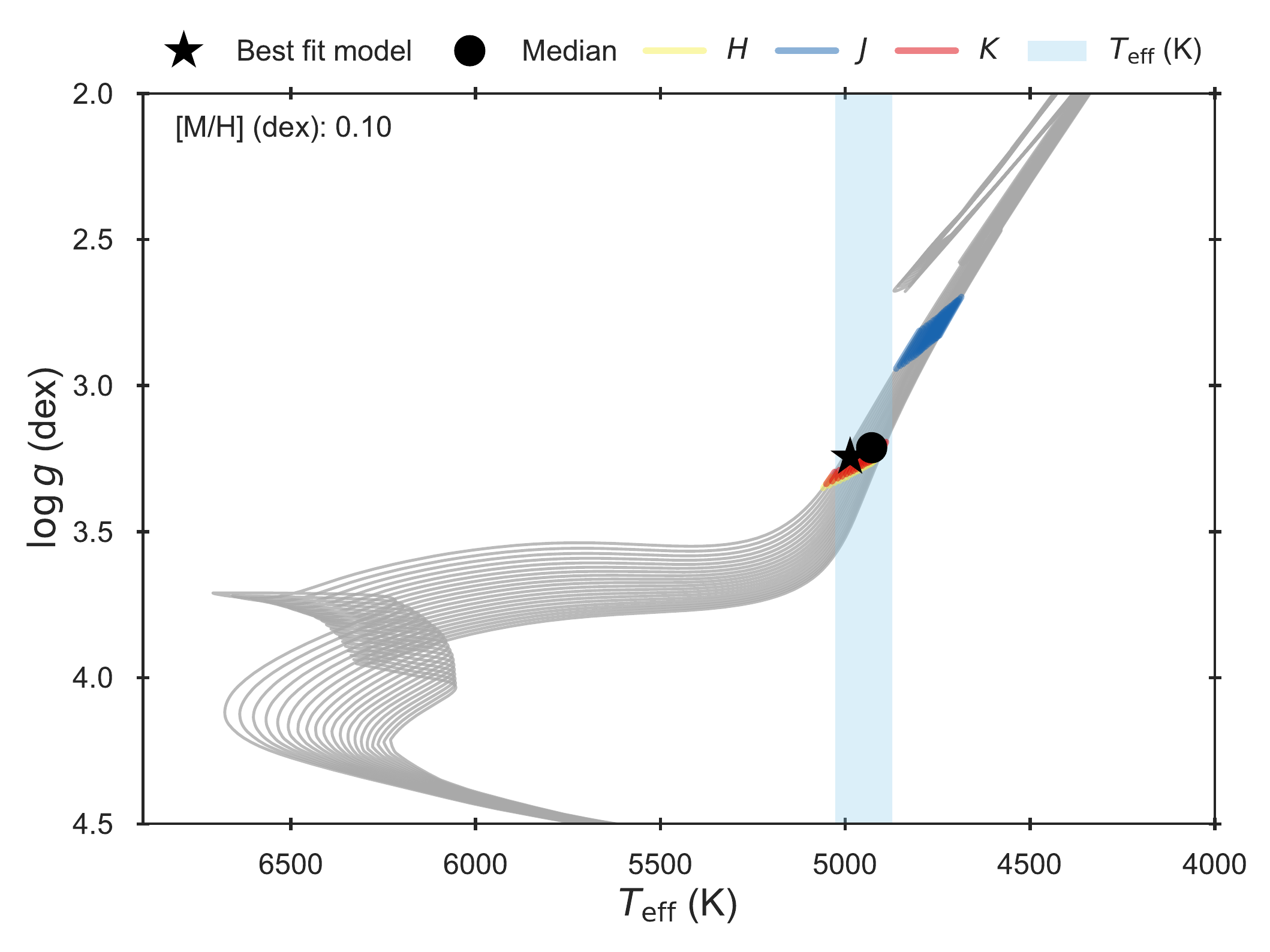}
    \includegraphics[width=84mm]{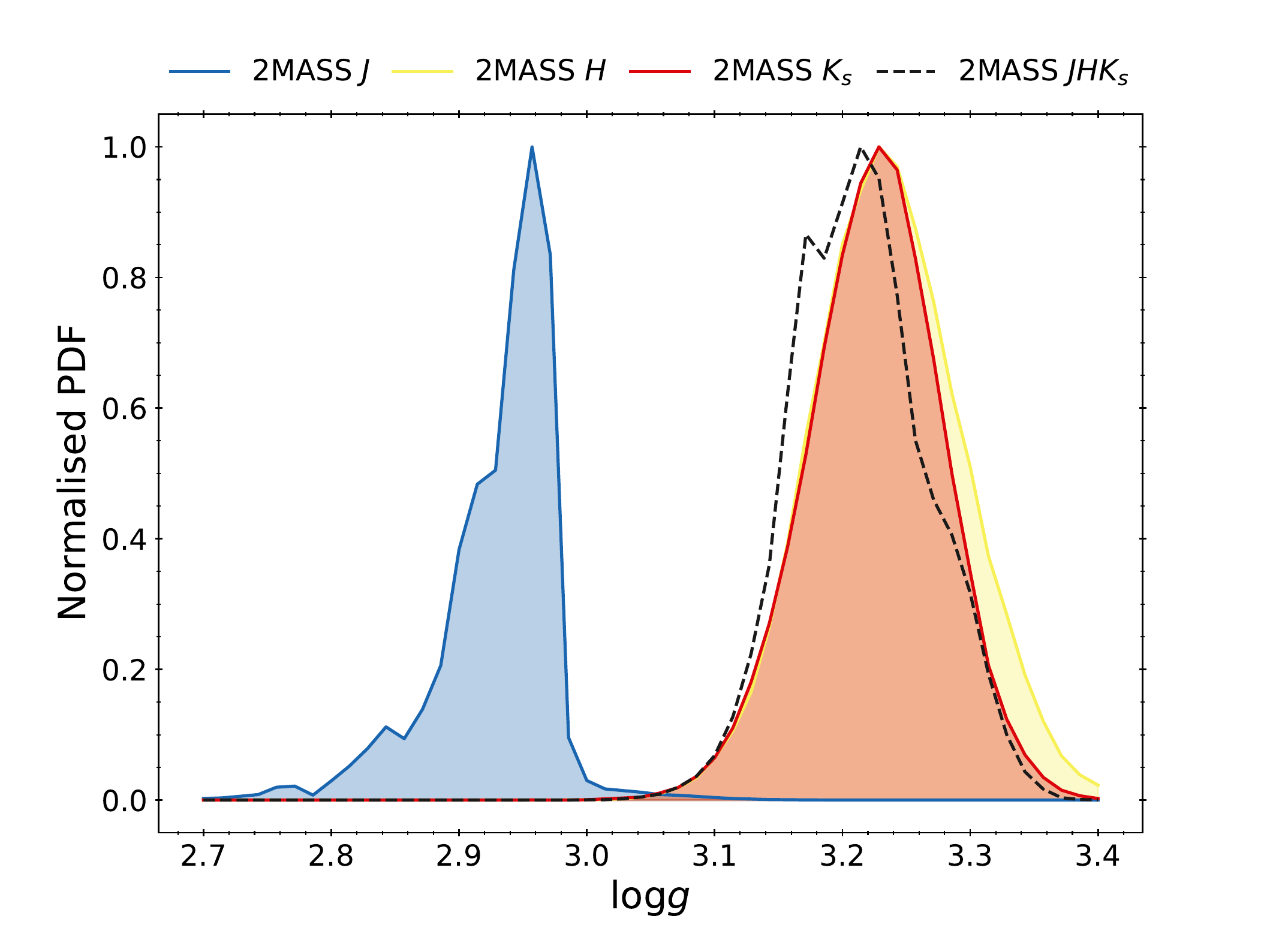}
    \caption{{\it Left:} Kiel diagram depicting BaSTI isochrones color-coded according to their correspondence to each fitted parameter. {\it Right:} Resulting $\logg$ probability density function as predicted in each filter, as well as the joint distribution obtained from Eq.~\ref{eq:likelihood} (dashed line). See text for details.}
    \label{fig:radius_magnitude_off}
\end{figure*}

Figure~\ref{fig:radius_magnitude_off} shows an example of the distance sampling procedure. The fitting includes effective temperature, metallicity, and parallax using the 2MASS filters. For the purpose of this example we have modified the apparent $J$-magnitude to be in disagreement with the other two magnitudes, as it can be seen in the Kiel diagram. The right panel of Fig.~\ref{fig:radius_magnitude_off} shows the probability density function of $\logg$ predicted by each of the filters. If BASTA did not sample the tails of the magnitude distributions far from their median values and standard deviations, the full likelihood as defined in Eq.~\ref{eq:likelihood} will be zero. The designed sampling scheme avoids this singularity and provides a robust statistical solution.

In addition to fitting parallax directly, distances can be determined in \BASTA independent of any parallax input as long as one photometric apparent magnitude is provided to the code. In this case, we solve for the distance and absorption iteratively using the magnitudes and the dust map until convergence is reached (normally within 3 iterations). If multiple apparent magnitudes are given as input we derive a distance in each passband and determine a joint distance by multiplying the individual probability distribution functions.
\section{Interpolation}\label{sec:interp}
\begin{figure*}
    \centering
    \includegraphics[width=\linewidth]{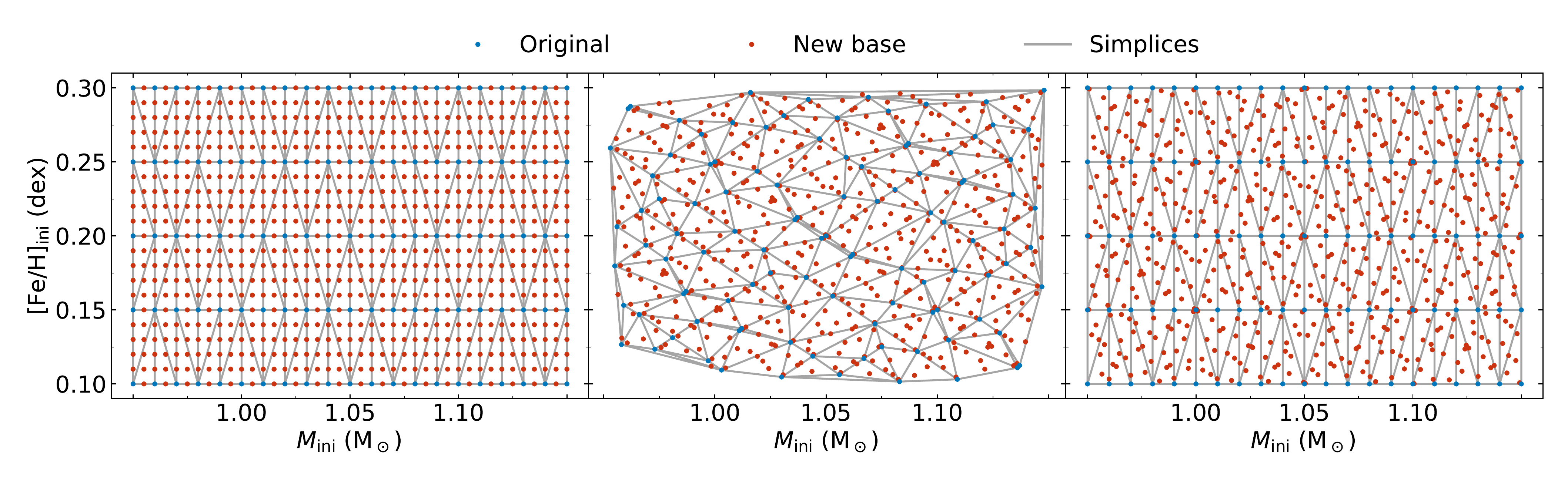}
    \caption{Example of how new tracks are assigned for interpolation across tracks for the case of a base consisting of $M_\mathrm{ini}$ and $\feh_\mathrm{ini}$. The simplices show the tessellation of the original grid, where each new track is interpolated to by using the tracks forming the simplex it is contained within. The left panel shows the assignment of tracks for a Cartesian input grid with a starting resolution of $0.05 \,\mathrm{dex}$ in initial metallicity and $0.01\,\msun$ in initial mass, and a desired interpolated resolution of $0.01\,\mathrm{dex}$ and $0.005\,\msun$ respectively. The middle panel shows the assignment of tracks for a Sobol input grid with an increase in the number of tracks by a factor of 2. The right panel shows the assignment of tracks for a Cartesian input grid but with Sobol assignment of tracks with an increase in resolution by a factor of 2. See text for details.}
    \label{fig:interpolation_base}
\end{figure*}
If the resolution of the grid used with \BASTA is lower than what is desired, the grid can be interpolated to match a user-defined resolution. \BASTA includes this option and can interpolate along or across stellar evolution tracks or isochrones (or both along and across, which we refer to as the combined option), and every parameter of the grid can be included in this interpolation. The individual procedures for each type of interpolation are described in the following sections.

To minimize computation time, the interpolation of tracks/isochrones can be limited to only a section of the original grid given by limits in any of the grid parameters. When fitting multiple stars simultaneously, these limits can be applied on a star-by-star basis (and thus producing one interpolated sub-grid per star), or as a single set of limitations for the full collection of stars (producing only one new sub-grid of tracks/isochrones).
\subsection{Interpolation along tracks/isochrones}\label{ssec:intpol_along}
For interpolation along the tracks/isochrones the user must define two relevant quantities. The first is the desired resolution in a stellar property between consecutive points in the track, normally an observed quantity that will be fitted (e.g., large frequency separation, or individual oscillation frequencies). The second relevant quantity is a (smoothly varying) grid parameter to be used as the independent variable in the interpolation. We refer to this parameter as the "base parameter", where typical examples are age, central hydrogen content, or central density for stellar evolution tracks, or the initial mass for the case of isochrones. The user can choose which base parameter will be used, or \BASTA will consider age/initial mass as default for stellar tracks/isochrones.

Once these two quantities are defined, the number of points along each interpolated track/isochrone is estimated as the number of points needed to satisfy the desired resolution assuming an equal spacing in the base parameter. The interpolation is performed as a 1-dimensional function using either a linear or cubic method via \texttt{scipy.interpolate.interp1d} \citep[from the \texttt{scipy} package,][]{scipy} on a track-by-track basis. The new base along with the interpolated parameters are stored and replace the original track. Therefore one should consider these as completely new evolutionary tracks, instead of simple refinements of the original.
\subsection{Interpolation across tracks/isochrones}\label{ssec:intpol_across}
\begin{figure*}
    \centering
    \includegraphics[width=\linewidth]{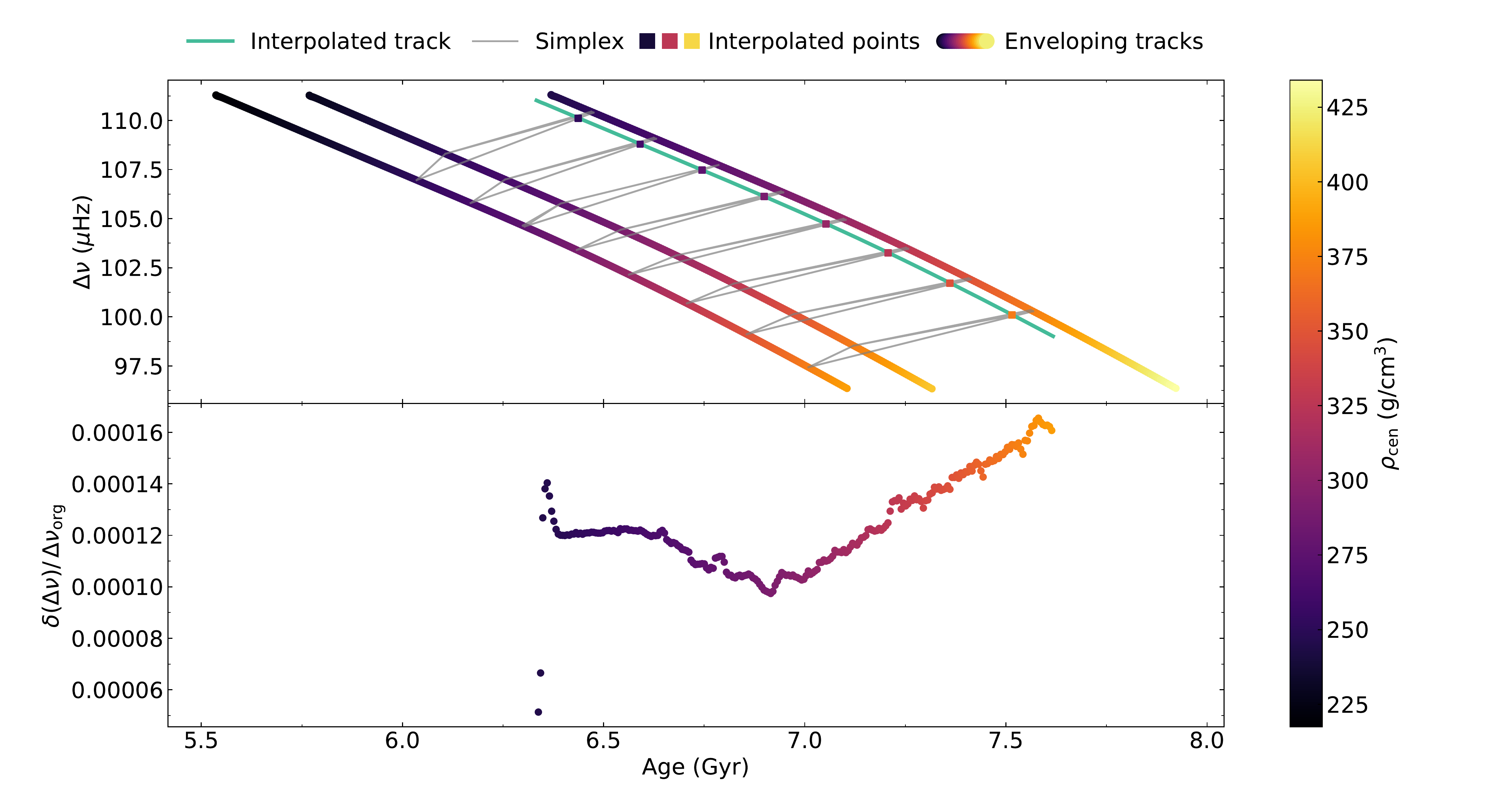}
    \caption{Example of the performance of our interpolation scheme. {\it Top:} large frequency separation as a function of age for the interpolated track (mint green) and the enveloping tracks used for the interpolation. The latter have been color-coded according to the base quantity used along the track (the central density $\rho_\mathrm{cen}$). {\it Bottom:} fractional difference in $\dnu$ between the original and interpolated track. See text for details.}
    \label{fig:interpolation_comparison}
\end{figure*}
Interpolation across tracks or isochrones can be applied to grids of identical microphysics. The base for this interpolation is formed by the parameters used to construct the original grid (see Section~\ref{sec:grids}), and their spacing gives the original grid resolution. For a Cartesian interpolated grid a desired resolution can be set for each of these parameters, and the code determines the minimal amount of tracks needed to satisfy this resolution one parameter at the time. For a Sobol-sampled interpolated grid the user defines a multiplicative increase in the number of tracks of the selected section of the grid. \BASTA then automatically determines a homogeneous distribution of new tracks in the base parameters that meets the required increase in the number of tracks. To retain this homogeneity, the resulting grid consists solely of interpolated tracks and the original ones are excluded. An example of these tracks assignments can be seen in Fig.~\ref{fig:interpolation_base}. We note that a Sobol sampled interpolated grid can be constructed starting from a Cartesian grid, as shown in the right panel of the figure.

The enveloping tracks used to interpolate each new track are determined from a tessellation of the base parameters using \texttt{scipy.spatial.Delaunay} (based on the \textsc{qhull} package \cite{10.1145/235815.235821}). The number of tracks considered to envelop a new track correspond to the number of parameters in the base plus one. In the example shown in Fig.~\ref{fig:interpolation_base} this requires three enveloping tracks for each new track (there are two parameters in the base, $M_{\rm ini}$ and $[{\rm Fe}/{\rm H}]_{\rm ini}$). The tessellation divides the original grid into triangles, as it can be seen in the example.

In addition to the enveloping tracks, the user must define a single additional smoothly varying quantity that runs along the evolution of the track to perform the interpolation (e.g., age, central hydrogen content, central density). We note that, when applying the combined interpolation method, this quantity does not need to be the same as the one selected for interpolation along the track. For example, the user can choose central hydrogen content when interpolating across tracks, and the subsequent interpolation along tracks can be performed with age as the independent variable.

To avoid extrapolations, the range of the smoothly varying parameter in the new track is limited to an interval that is contained by all enveloping tracks, and its spacing is determined as the mean of the spacing in the enveloping tracks. Using this basis, each parameter in the new track is interpolated separately using \texttt{scipy.interpolate.LinearNDInterpolater} (that relies on a new tessellation of the points along each enveloping track). Individual oscillation frequencies are treated as independent parameters, and are therefore interpolated individually.

As a test of our interpolation procedure, Fig.~\ref{fig:interpolation_comparison} shows the results of reconstructing an evolutionary track extracted from the Sobol grid presented in the middle panel of Fig.~\ref{fig:interpolation_base}. The top panel depicts the evolution of the large frequency separation $\dnu$ as a function of age for the three enveloping tracks determined from the tessellation of the base parameters, as well as the track reconstructed with our combined interpolation approach. The bottom panel presents the outcome of the combined interpolation procedure as the fractional difference in $\dnu$ between the interpolated track and the original track. Deviations are at the $10^{-4}$ level, an order of magnitude smaller than the uncertainty in $\dnu$ measured for the best \Kepler targets \citep[see Fig.~6 in][]{2018ApJS..236...42Y}.

\begin{figure*}
%    \centering
    \includegraphics[width=84mm]{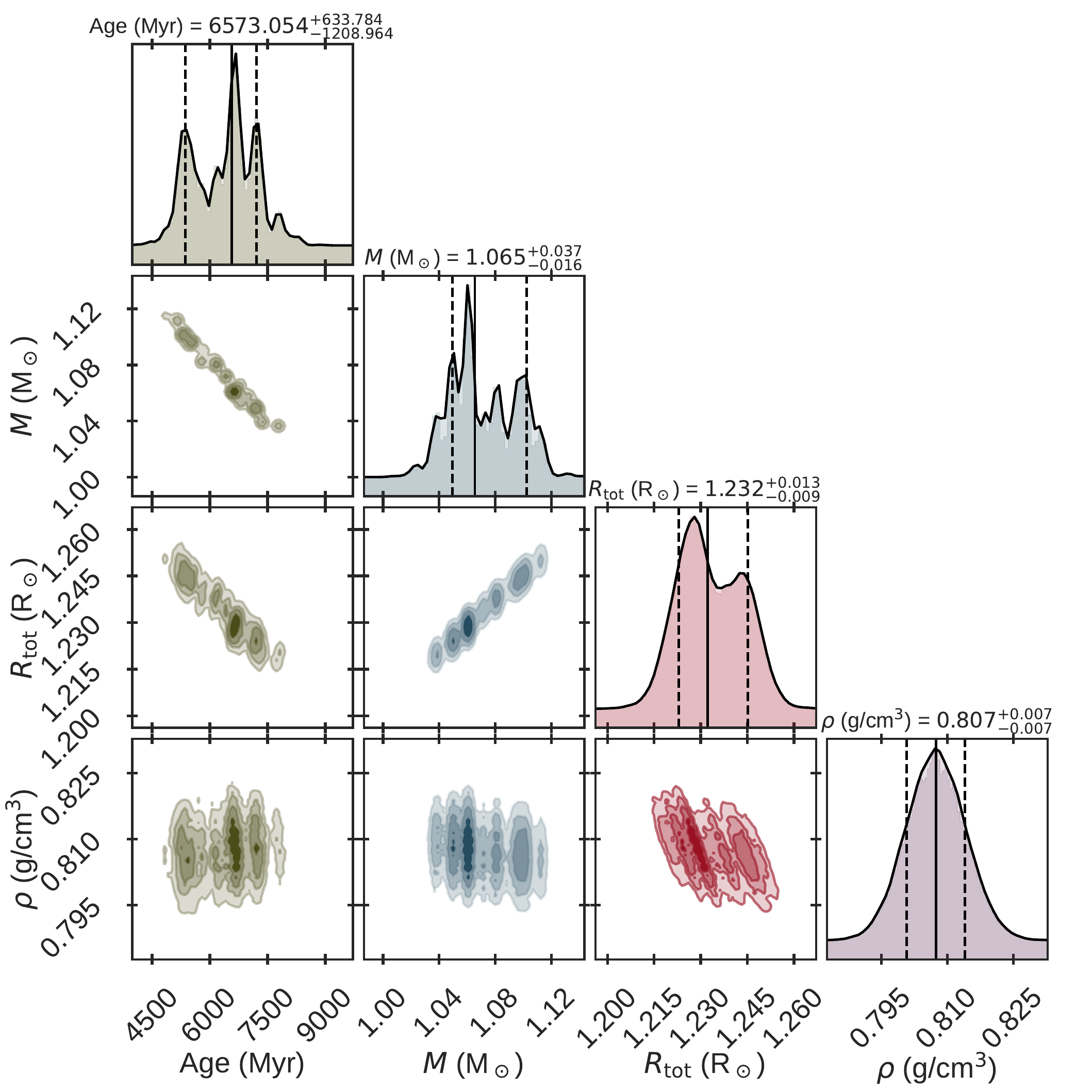}
    \includegraphics[width=84mm]{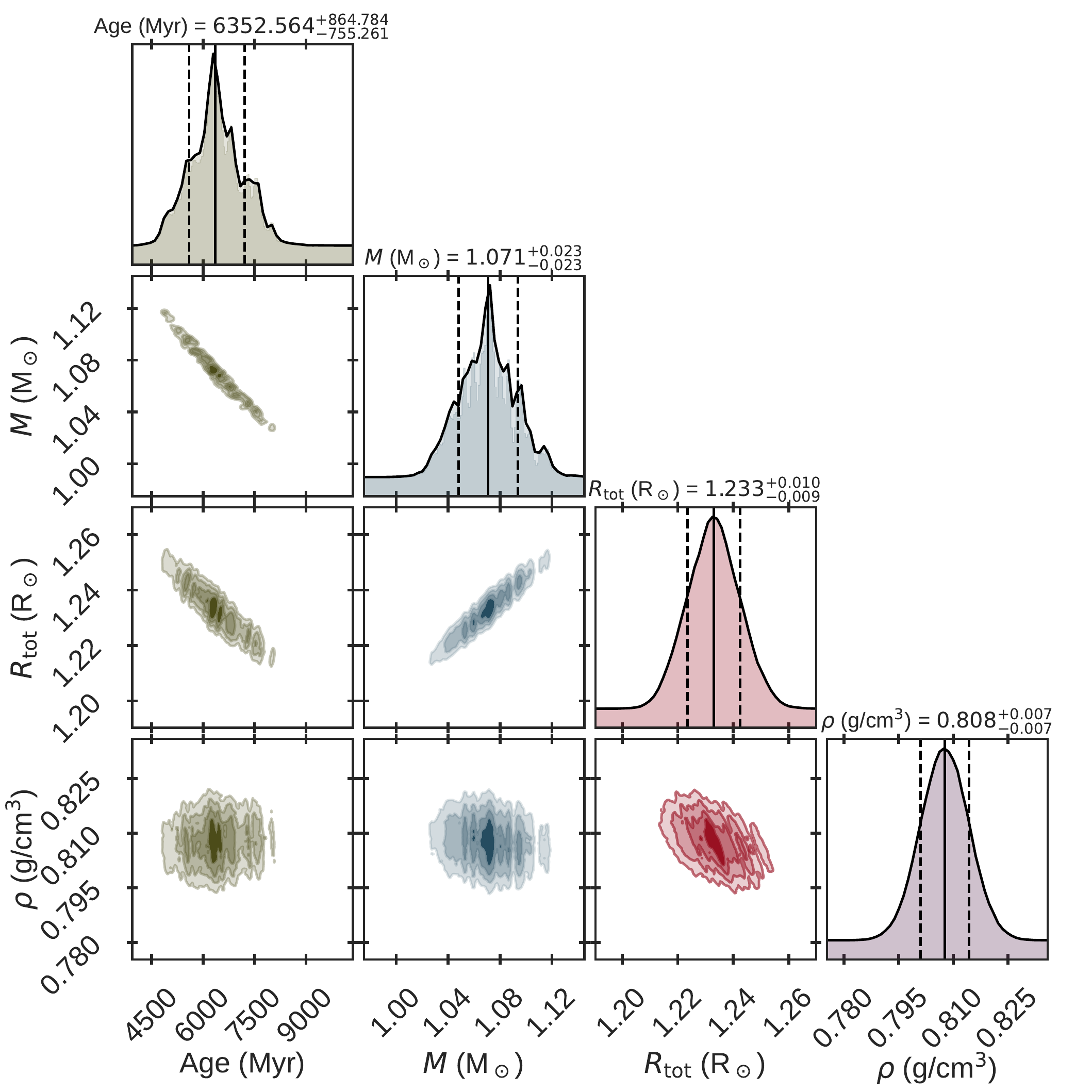}
    \caption{Posterior distributions of stellar parameters for the \Kepler target 16~Cyg~A obtained fitting the set ($\teff$, $\feh$, $\dnu$, $\num$) with a Sobol grid of different resolutions. {\it Left:} original grid. {\it Right:} interpolated grid. See text for details.}
    \label{fig:16CygA_corner}
\end{figure*}

The impact of our interpolation procedure can be seen in Fig.~\ref{fig:16CygA_corner}, where we compare the obtained posterior PDFs for mass, radius, age and density when fitting data of the \Kepler target 16~Cyg~A. For this example we considered as input parameter the effective temperature and metallicity from \citet{2009A&A...508L..17R} and the large frequency separation and frequency of maximum power derived by \citet{2017ApJ...835..172L}. These data were then fitted to a Sobol grid in its original resolution, and an interpolated grid constructed with the combined method and an increase in resolution by a factor of 5 in the number of tracks (across tracks) and spacing of $0.1\mu$Hz between the lowest observed $\ell=0$ frequency (along tracks). The derived quantities are in very good agreement (and certainly within their respective uncertainties), but the resulting distributions are significantly smoother after the interpolation procedure.

Before closing this section we note that when the combined method is chosen the interpolation across tracks/isochrones is performed before interpolation along the tracks/isochrones to increase computation efficiency. We have tested the inverse case (along before across), and confirmed the differences in the recovery procedure presented in Fig.~\ref{fig:interpolation_comparison} are of the same magnitude regardless of the order of the interpolation. This inverse case (along before across) is still available for usage in \BASTA at a much larger computational cost.
\section{Validation with artificial data}\label{sec:tests}
We performed a thorough end-to-end validation of \BASTA to quantify the robustness of our pipeline in retrieving stellar properties. For this purpose we produced artificial data from models extracted from a grid and used the same grid to fit the observables, which is equivalent to assuming that the underlying stellar models are true representations of the observations. By quantifying the deviations in our derived parameters from the true values we can estimate the level of accuracy of our pipeline, as these will depend exclusively in the reliability of \BASTA. In addition, our obtained uncertainties in stellar properties provide an estimate of the typical statistical precision, which in turn depends on the assumed observational errors and the combination of input quantities fitted. We emphasise that this exercise allows us to test the accuracy and statistical precision obtained for a given set of observables and their measured uncertainty, but we cannot account for deviations between the physical and observed properties of a real star and those predicted by a grid of stellar models. This additional systematic uncertainty undoubtedly exists and remains to be quantified, but it depends on our many shortcomings in theory of stellar evolution and goes well beyond the scope of this paper.

For this particular exercise we constructed a Sobol grid of stellar models using {\tt GARSTEC} comprising 3,000 evolutionary tracks, a mass range between $0.8-1.5$~M$_\odot$, initial metallicity $-0.5<$[Fe/H]$_\mathrm{ini}<+0.5$, the \citet{2009ARA&A..47..481A} solar mixture, and an enrichment law $\Delta Y/\Delta Z=1.4$. The tracks were evolved from the pre-main sequence to a value of the large frequency separation $\dnu=10~\mu$Hz, roughly corresponding to the lower RGB. We determined individual oscillation frequencies of angular degree $\ell=0,1,2,3$ for stars down to a value of $\dnu=30~\mu$Hz, and only radial modes for more evolved stars. The coverage of the grid was selected to ensure that we can test our fitting procedures in various types of stars (main sequence, subgiants, and red giants) which have different observed quantities to be fitted (individual frequencies, mixed-modes, period spacing).

We assume the following observational uncertainties (taken from \citet{2017ApJS..233...23S} and \citet{2017ApJ...835..172L} for main-sequence and subgiant targets, and \citet{2018ApJS..236...42Y} for red giant stars): 70~K in $\teff$, 0.1~dex in $\feh$, 0.1~dex in $\logg$, 0.5\% of the observed value in $\dnu$, and 2\% in $\num$. These are typical uncertainties in the global asteroseismic properties for \Kepler stars observed for more than 50 days \citep[see][]{2017ApJS..233...23S}. The asymptotic period spacing uncertainty is assumed to be 1.5~\%, which corresponds to the average uncertainty measured by \citet{2016A&A...588A..87V} for RGB stars in the range $10~\mu$Hz$\leq\dnu\leq30~\mu$Hz. For the individual frequencies of oscillation we adopt a two-step process where we first identify which modes would be detected given an assumed observation length, and then determine their uncertainties following a recipe derived from \Kepler targets.
\begin{figure*}
%\centering
\includegraphics[width=84mm]{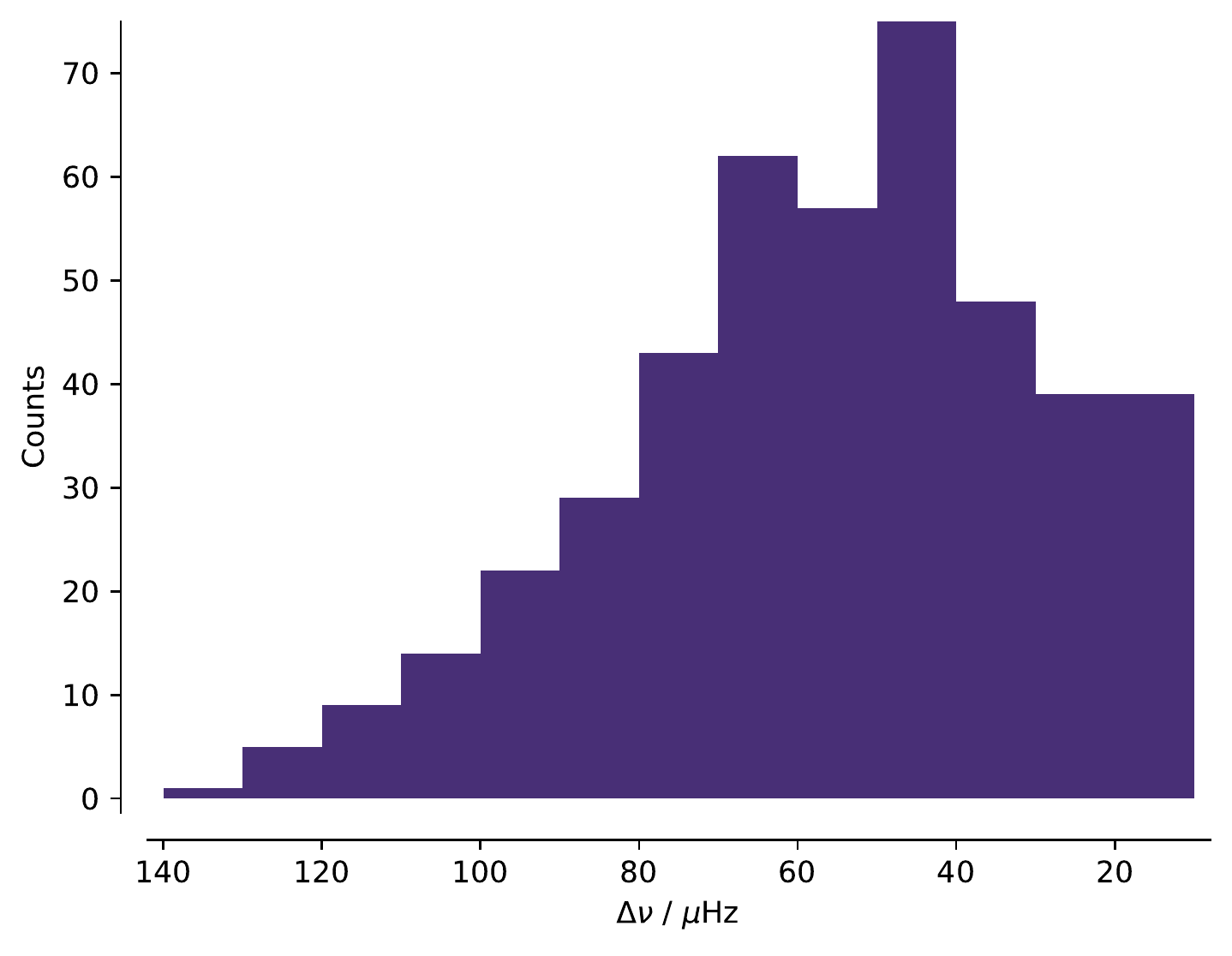}
\includegraphics[width=84mm]{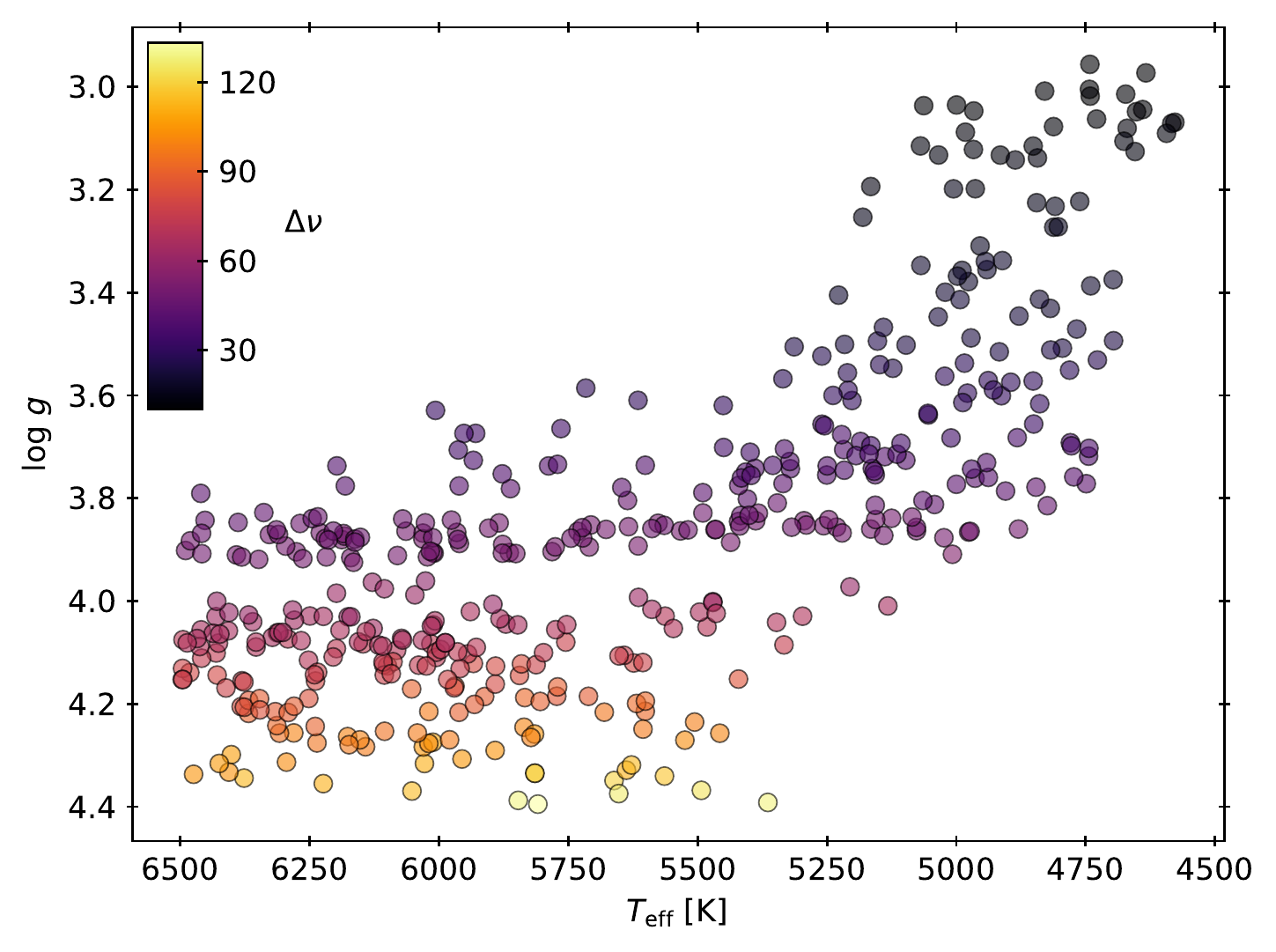}
\caption{{\it Left:} Distribution in large frequency separation of the synthetic data used for the validation procedure. {\it Right:} Kiel diagram depicting the distribution of our synthetic targets across the grid, color-coded by the value of the large frequency separation.}
\label{fig:Kielval}
\end{figure*}

For a given theoretical mode calculation we estimate which modes would likely be measurable based on results from the main-sequence stars included in the LEGACY study \citep[][]{2017ApJ...835..172L}. From the LEGACY data we first estimate the typical minimum and maximum frequencies of measured $\ell=0$ modes in units of $\num$, and derive simple linear relations of these frequencies as a function of $\num$. The typical frequency intervals of measurable $\ell=0$ modes are found to range from $\pm0.4\, \num$ at $\num=\rm 1000\, \mu Hz$ to approximately $\pm0.2\,\num$ at $\num=\rm 4000\, \mu Hz$. We then estimate the expected relative amplitude of $\ell=0$ modes at these frequency limits, using a relation for the envelope width of the assumed Gaussian modulation of mode amplitudes around $\num$. This relation is determined from fits to LEGACY data, and includes both a dependence on $\num$ and $T_{\rm eff}$ (Lund et al., in prep.). By including mode visibilities from \citet[][]{2017ApJ...835..172L} we can then estimate the corresponding relative amplitudes of non-radial modes and assess which of these exceeds the limit for detectability set by the $\ell=0$ modes. Following \citet[][]{ball2018} we include information on the mode inertia for non-radial modes, where we divide amplitudes by the square-root of the $Q$-factor (ratio of the mode inertia relative to the $\ell=0$ inertia at the corresponding frequency, see also Eq.~\ref{eqn:k08})

Concerning frequency uncertainties for the modes deemed measurable from the above procedure we use a polynomial relation between the frequency uncertainties in units of $\num$ of $\ell=0$ modes from the LEGACY data and the corresponding relative frequency away from $\num$. At $\num$ the typical minimal uncertainty is found to be of the order ${\sim}7.6\times 10^{-5}\, \num$, which for a $\num=\rm 2000\, \mu Hz$ star corresponds to $\rm \sigma_{\nu}\sim0.15\, \mu Hz$. Away from $\num$ the typical uncertainties increase by factors of $5$ ($\nu \sim -0.4\, \num$) to $10$ ($\nu \sim +0.4\, \num$).

With the observational uncertainties in all relevant observational quantities defined, the validation procedure was designed as follows. We 
selected 443 models from the grid to be used as artificial targets mimicking the observed distribution in large frequency separation of the \Kepler main-sequence and subgiant sample of \citet{2017ApJ...835..172L} (see Fig~\ref{fig:Kielval}). This sample reaches values of $\dnu\simeq20\mu$Hz, which we extended in same proportion as the last bin to $\dnu\simeq10\mu$Hz to encompass the base of the RGB. We considered various sets of input quantities (see Table~\ref{tab:summHH}) and generated the synthetic data by drawing a sample from a normal Gaussian distribution with the model value of the relevant parameter as the mean and a standard deviation given by the observational uncertainty described above. In that manner, each quantity in the input set does not exactly correspond to the model value but it has been perturbed by typical observational uncertainties. This is the closest we can simulate observations of a real star, under the assumption that the underlying grid of models is a true representation of the physics at play in stellar evolution.
\begin{figure*}
\centering
\includegraphics[width=\textwidth]{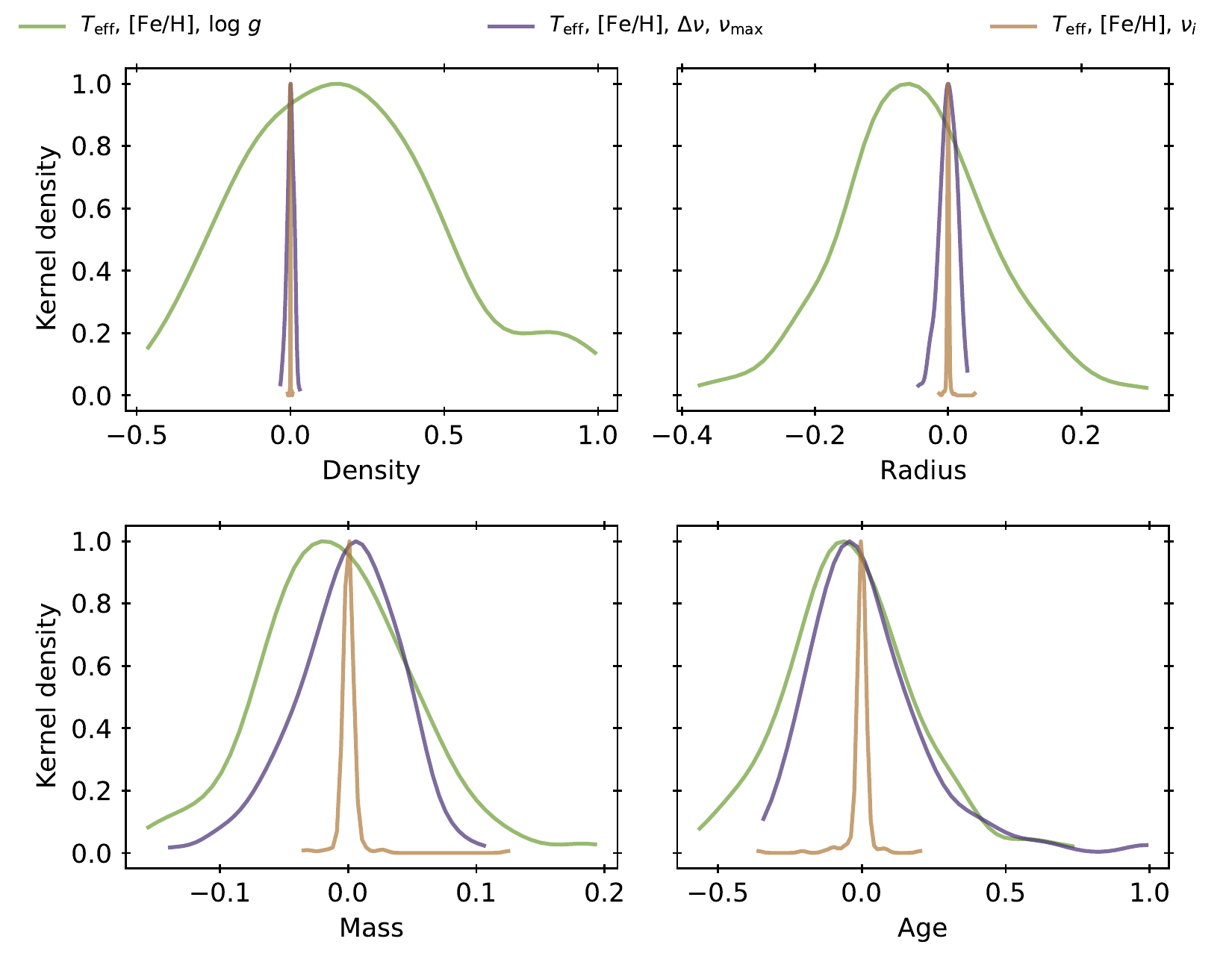}
\caption{Distribution of normalised differences between derived stellar properties with \BASTA and the underlying stellar model for the main-sequence targets. Values are computed as (Solution-Model)/Model. See text for details.}
\label{fig:HHpdfsMS}
\end{figure*}

The distribution of the recovered fractional values in density, radius, mass, and age for the main-sequence artificial targets are shown in Fig.~\ref{fig:HHpdfsMS} (equivalent figures for the subgiant and RGB targets are given in Appendix~\ref{app:validation}). As we can see, the inferred properties are generally in good agreement with the corresponding underlying input parameters. The inclusion of asteroseismic data dramatically improves the quality of the recovered properties compared to the cases where only atmospheric parameters are used.

As is also visible in Fig.~\ref{fig:HHpdfsMS}, the inferred properties do not match exactly with the underlying model values. To ensure that this is a result of finite precision of the data, we reduced the assumed standard uncertainties by a factor of $10^{-4}$ and repeated the above validation. \BASTA recovered the original model in 440 out of 443 cases, and in 3 cases we pick the model just before or after which is expected due to the perturbation to the model parameters that defines the input observables. These results confirm that the differences are due to finite data precision and are not an artifact of combined effects of priors and weights.

Since we have determined stellar properties assuming typical observed uncertainties in various sets of input, we can use the \BASTA results to quantify the statistical uncertainty obtained across evolutionary phases for various sets of input. Table~\ref{tab:summHH} summarizes the results, where we have listed for each stellar property the average of the 16th and 84th percentiles normalized by the median derived by \BASTA. It is clear from the table how the accuracy increases as asteroseismic quantities are included in comparison to just the spectroscopic input, as well as the additional gain from including individual frequencies. This exercise also shows that, if our stellar models are a good representation of the observed stars, statistical uncertainties below the 10\% level in age are within reach thanks to the advent of asteroseismology. 

Another interesting point arising from the validation is the extremely high precision in the recovered stellar properties for subgiant stars when individual oscillation frequencies are fitted. As mixed-modes evolve rapidly and are very sensitive to the conditions in the stellar core, such a small statistical uncertainty is not surprising as long as the code is capable of identifying the correct underlying model. We remind the reader that in subgiant stars the number of model non-radial oscillation frequencies between two radial modes exceeds the number of observed frequencies as their visibility depends in the inertia. The procedure we described in Section~\ref{sssec:smixed} can successfully identify which mixed-modes should be visible and thus avoids an artificial increase in the statistical uncertainties due to miss-identified modes. We include a few representative \'Echelle diagrams from the validation procedure in Appendix~\ref{app:validation} where the accuracy of the algorithm is demonstrated.
\begin{table*}
\renewcommand{\arraystretch}{1.2}
\centering
\caption{Fractional uncertainties in stellar properties across evolutionary phases determined for different sets of input. The values are calculated as the average of the 16th and 84th percentiles normalised by the derived median. See text for details.}
\label{tab:summHH}
\begin{tabular}{l l l c c c c}
\hline
Phase & Input set & & $\delta\rho/\rho$ & $\delta R/R$ & $\delta M/M$ & $\delta$Age$/$Age \\
\hline
MS & {$\teff$, $\feh$, $\logg$} & & $0.3223$ & $0.1179$ & $0.0558$ & $0.2880$ \\
$\dnu\ge60$ & {$\teff$, $\feh$, $\dnu$, $\num$} & & $0.0101$ & $0.0124$ & $0.0364$ & $0.1620$ \\
(185 stars) & {$\teff$, $\feh$, $\nu_i$} & & $0.0005$ & $0.0010$ & $0.0029$ & $0.0177$ \\
\hline
SG & {$\teff$, $\feh$, $\logg$} & & $0.3290$ & $0.1356$ & $0.0947$ & $0.3262$ \\
$60<\dnu<30$ & {$\teff$, $\feh$, $\dnu$, $\num$} & & $0.0098$ & $0.0142$ & $0.0393$ & $0.1175$ \\
(178 stars) & {$\teff$, $\feh$, $\nu_i$} & & $0.0001$ & $0.0002$ & $0.0005$ & $0.0009$ \\
\hline
RGB & {$\teff$, $\feh$, $\logg$} & & $ 0.3240$ & $0.1283$ & $0.1291$ & $0.4635$ \\
$\dnu\le30$ & {$\teff$, $\feh$, $\dnu$, $\num$} & & $0.0101$ & $0.0213$ & $0.0592$ & $0.2227$ \\
(80 stars) & {$\teff$, $\feh$, $\nu_i$} & & $0.0009$ & $0.0075$ & $0.0217$ & $0.0937$ \\
& {$\teff$, $\feh$, $\dnu{}$, $\Delta$P$_1$} & & $0.0097$ & $0.0313$ & $0.0915$ & $0.3447$ \\
\hline
Full Sample & {$\teff$, $\feh$, $\logg$} & & $0.3253$ & $0.1269$ & $0.0847$ & $0.3350$ \\
& {$\teff$, $\feh$, $\dnu$, $\num$} & & $0.0099$ & $0.0147$ & $0.0417$ & $0.1551$ \\
& {$\teff$, $\feh$, $\nu_i$} & & $0.0004$ & $0.0018$ & $0.0053$ & $0.0247$ \\
\hline
\end{tabular}
\end{table*}
\section{Notes about code development and contributing}\label{sec:codedev}
The core developers of \BASTA have customized the code following the needs of our own research fields. As the code begins to be used by scientists around the world, we expect that the inclusion of additional features will become highly desirable. Naturally the amount of proposed improvements to the code will increase proportionally to the number of users, and it will go beyond the available time of the core developers to implement them and maintain a stable version of \BASTA.

For these reasons we rely on the {\tt git} version-control system and the {\tt GitHub} repository\footnote{\url{https://github.com/BASTAcode/BASTA}} to encourage users to develop and share their contributions with the rest of the users. New features, additions, or extensions should be in separate branches/forks so that they do not need the direct involvement of the core developing team. Once the new additions are completed, the code developers will be happy to handle the pull request and make it part of the stable version of \BASTA after proper testing has been completed. A more detailed contribution guide can be found on the {\tt GitHub} repository along with the installation instructions and a tracker for issues/bugs/suggestions. We encourage the users to follow the project on {\tt GitHub} (please ``watch'' and select ``custom/releases'' as a minimum) to get notifications and updates on new developments. We welcome any participation including pull requests.
\section{Concluding remarks}\label{sec:conc}
We have introduced the public version of the BAyesian STellar Algorithm (\BASTA), an open-source code that determines stellar properties using a set of observables and a grid of stellar models/isochrones. It is flexible in its input and can combine a large number of spectroscopic, photometric, astrometric, and asteroseismic input to extract properties of stars under a statistically robust Bayesian scheme. The large number of functionalities included in \BASTA, combined with the various sets of publicly available or custom-computed grids of stellar models/isochrones, make it the most versatile pipeline for stellar analysis currently available.

We have thoroughly described the type of input that can be given to retrieve stellar properties, and discussed the assumptions made when predicting these quantities from the underlying grids of evolutionary models. We have performed an extensive validation test that confirms the reliability of \BASTA in determining accurate stellar properties, and use these results to quantify the typical statistical uncertainties obtained for various combinations of fitting parameters. Our results show that asteroseismic ages with statistical uncertainties below the 10\% level are achievable for datasets of the quality obtained by the \Kepler satellite, and the only limiting factor is the reliability of our stellar evolution and pulsation models.

We hope to have provided the community with a useful analysis tool for stellar properties, which is specifically designed to meet the challenges of large inhomogeneous datasets available in the era of large-scale stellar surveys. Its capability of combining a wide range of input data makes it an invaluable tool to fully exploit the potential of current stellar catalogues (by simultaneously fitting e.g., TESS, APOGEE, 2MASS, and Gaia eDR3 data), and its future developments will make it ready for the challenge of the next data deluge from surveys such as PLATO 2.0 \citep{2014ExA....38..249R}, WEAVE \citep{2014SPIE.9147E..0LD}, 4MOST \citep{2019Msngr.175....3D}, and the Legacy Survey of Space and Time at the Vera Rubin Observatory \citep[LSST,][]{2019ApJ...873..111I}.
\section*{Acknowledgements}
Funding for the Stellar Astrophysics Centre is provided by The Danish National Research Foundation (Grant agreement No.~DNRF106). VAB-K acknowledges support from VILLUM FONDEN (research grant 10118), the Independent Research Fund Denmark (Research grant 7027-00096B), and the Carlsberg foundation (grant agreement CF19-0649). SC acknowledges support from Premiale INAF MITiC (PI: B. Garilli), from Istituto Nazionale di Fisica Nucleare (INFN) (Iniziativa specifica TAsP), from Progetto Mainstream INAF (PI: S. Cassisi) and grant AYA2013- 42781P from the Ministry of Economy and Competitiveness of Spain. AMS is partially supported by grant PID2019-108709GB-I00 of the MICINN of Spain.
AS acknowledge support from the European Research Council Consolidator Grant funding scheme (project ASTEROCHRONOMETRY, G.A. n. 772293, \url{http://www.asterochronometry.eu}).

\section*{Data availability}
The data underlying this article will be shared on reasonable request to the corresponding author.\\

The following software was used during this research:
\textsc{astropy} \citep{astropy},
\textsc{corner} \citep{corner}, 
\textsc{dustmaps} \citep{2018JOSS....3..695M},
\textsc{emcee}  \citep{emcee},
\textsc{h5py} \citep{h5py}, 
\textsc{healpy} \citep{healpy},
\textsc{matplotlib} \citep{matplotlib},
\textsc{numpy} \citep{numpy},
\textsc{scikit-learn} \citep{scikit-learn}, 
\textsc{scipy} \citep{scipy}, and
\textsc{tqdm} \citep{tqdm}.


\bibliographystyle{mnras}
\bibliography{bastalib} % if your bibtex file is called example.bib

% %%%%%%%%%%%%%%%%% APPENDICES %%%%%%%%%%%%%%%%%%%%%
\appendix
\section{Photometric systems}\label{app:colors}
Table~\ref{tab:photcolors} gives a list of the currently available synthetic colors for all tracks and isochrones used with \BASTA. Additional sets are added as new passbands become available.
\begin{table*}
\centering
\caption{Available photometric systems.}
\label{tab:photcolors}
\begin{tabular}{l l l l}
\hline
\hline
 Photometric system & Calibration & Passbands & Zero-points \\
\hline
{\it UBVRIJHKLM} & Vegamag & \citet{1988PASP..100.1134B}; \citet{1990PASP..102.1181B} 
& \citet{Bessell98} \\
{\it HST}-WFPC2 & Vegamag & SYNPHOT & SYNPHOT \\
{\it HST}-WFC3 & Vegamag & SYNPHOT & SYNPHOT \\
{\it HST}-ACS & Vegamag & SYNPHOT & SYNPHOT \\
2MASS & Vegamag & \citet{2003AJ....126.1090C} & \citet{2003AJ....126.1090C} \\
DECam & ABmag & DES collaboration & 0 \\
{\it Gaia DR1} & Vegamag & \citet{Jordi2010} & \citet{Jordi2010} \\
{\it Gaia DR2} & Vegamag & \citet{gaiadr2} &   \citet{gaiadr2} \\
{\it Gaia eDR3} & Vegamag & \citet{gaiadr3} & \citet{gaiadr3} \\
{\it JWST}-NIRCam & Vegamag & {\it JWST} User Documentation & SYNPHOT \\
SAGE & ABmag & SAGE collaboration & 0 \\
Skymapper & ABmag & \citet{2011PASP..123..789B} & 0 \\
Sloan & ABmag & \citet{doi:2010} & \citet{2008ApJS..178...89D} \\
Str{\"o}mgren & Vegamag & \citet{2006AJ....131.1184M} & \citet{2006AJ....131.1184M} \\
VISTA & Vegamag & ESO & \citet{rubele:12} \\
Tycho$+$Hipparcos & ABmag  & \citet{bm:12}  & \citet{bm:12}  \\
{\it Kepler} & ABmag  & Kepler collaboration  & 0 \\
{\it TESS} & ABmag & TESS collaboration & 0  \\
{\it WISE} W1 \& W2 & Vegamag & WISE Collaboration & \citet{wise}\\
\hline
\end{tabular}
\end{table*}

\section{Validation figures for subgiant and RGB targets}\label{app:validation}
We include in this section the validation figures for the subgiant and RGB targets, equivalent to Fig.~\ref{fig:HHpdfsMS}. We also show a few representative \'Echelle diagrams obtained in the validation procedure showing the performance of the frequency matching algorithm described in Section~\ref{sssec:smixed} in the presence of mixed-modes.
\begin{figure*}
\centering
\includegraphics[width=\textwidth]{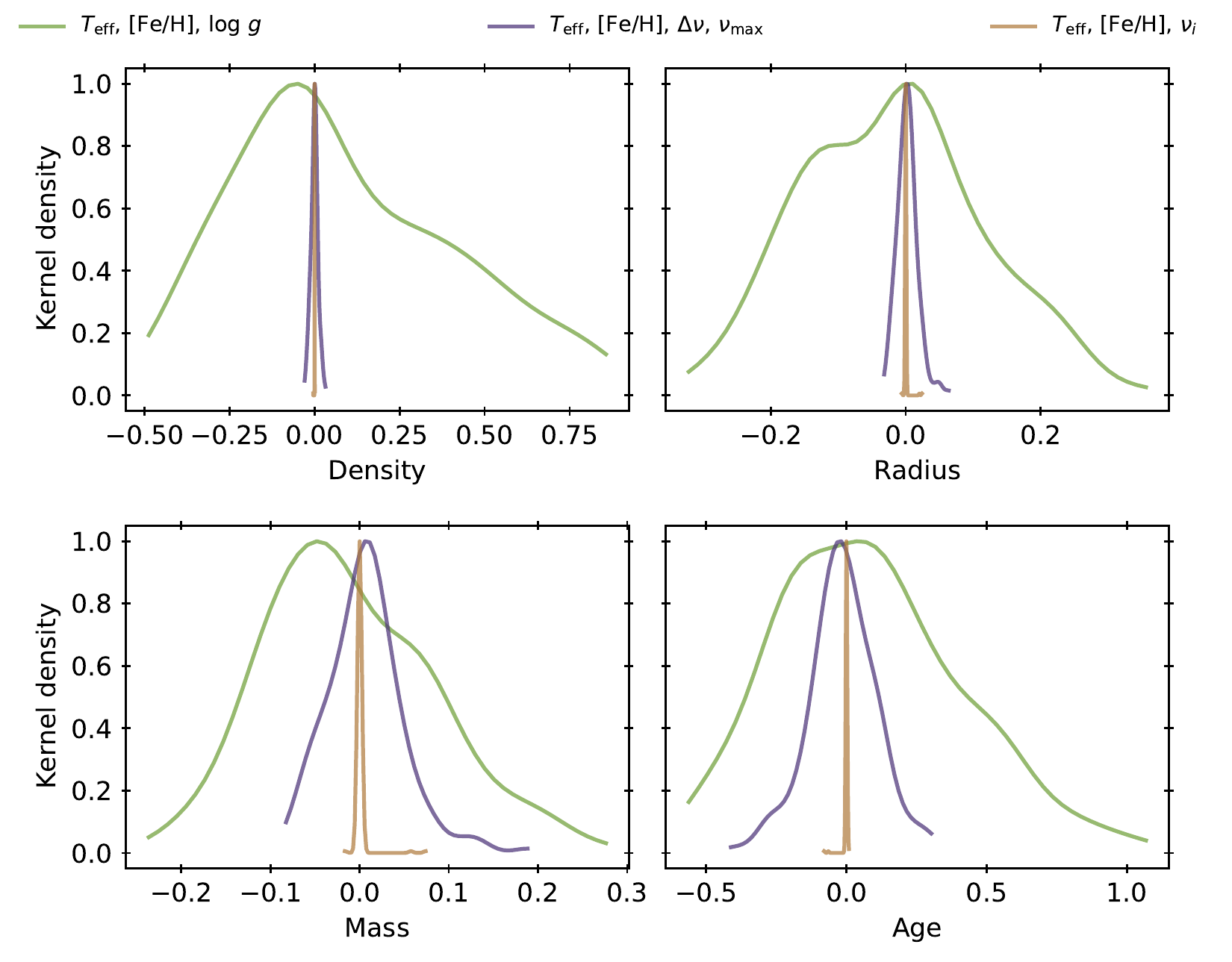}
\caption{Same as Fig.~\ref{fig:HHpdfsMS} for the targets in the subgiant phase.}
\label{fig:HHpdfsRG}
\end{figure*}
\begin{figure*}
\centering
\includegraphics[width=\textwidth]{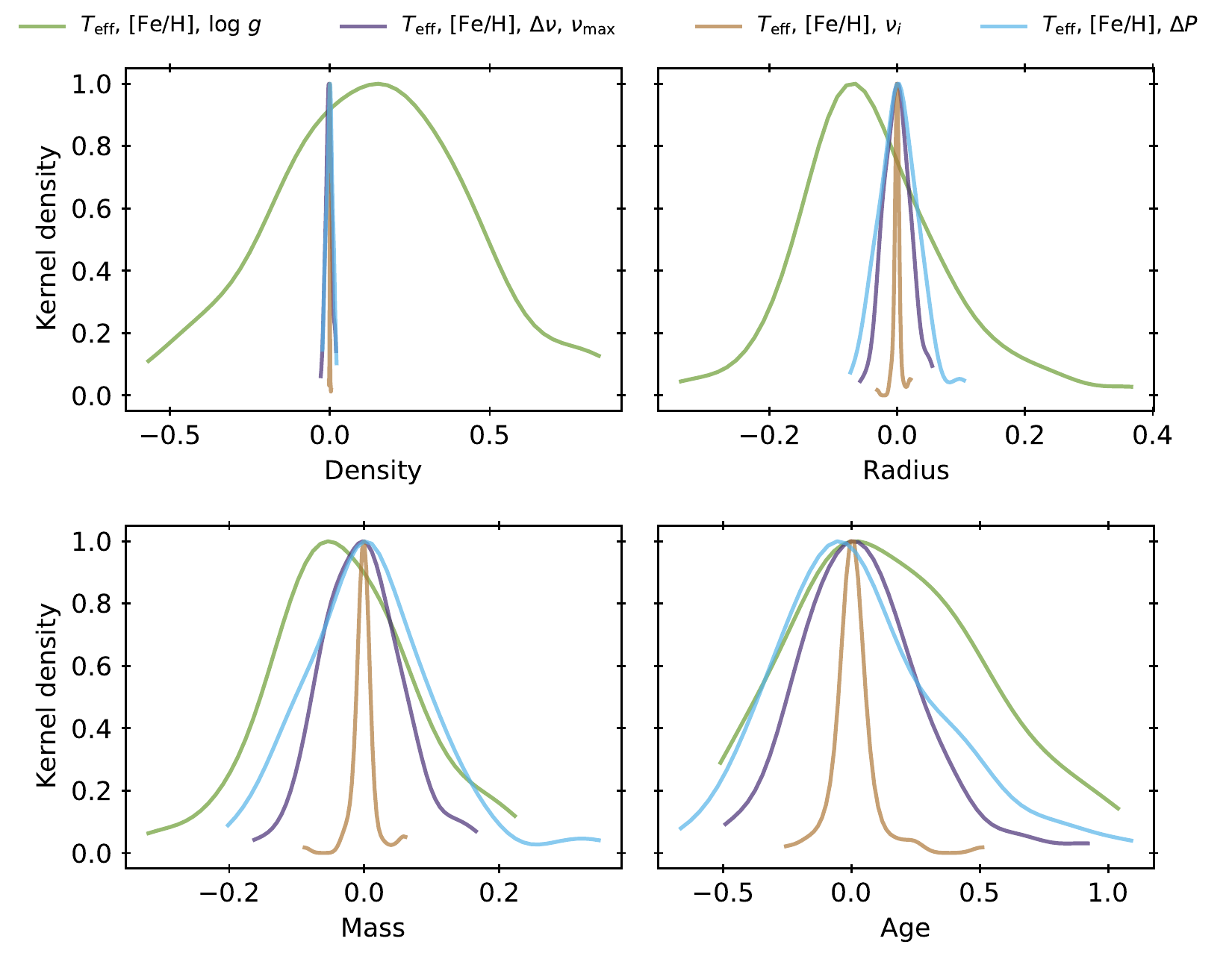}
\caption{Same as Fig.~\ref{fig:HHpdfsMS} for the targets in the RGB phase.}
\label{fig:HHpdfsRG}
\end{figure*}
\begin{figure}
\centering
\includegraphics[width=\linewidth]{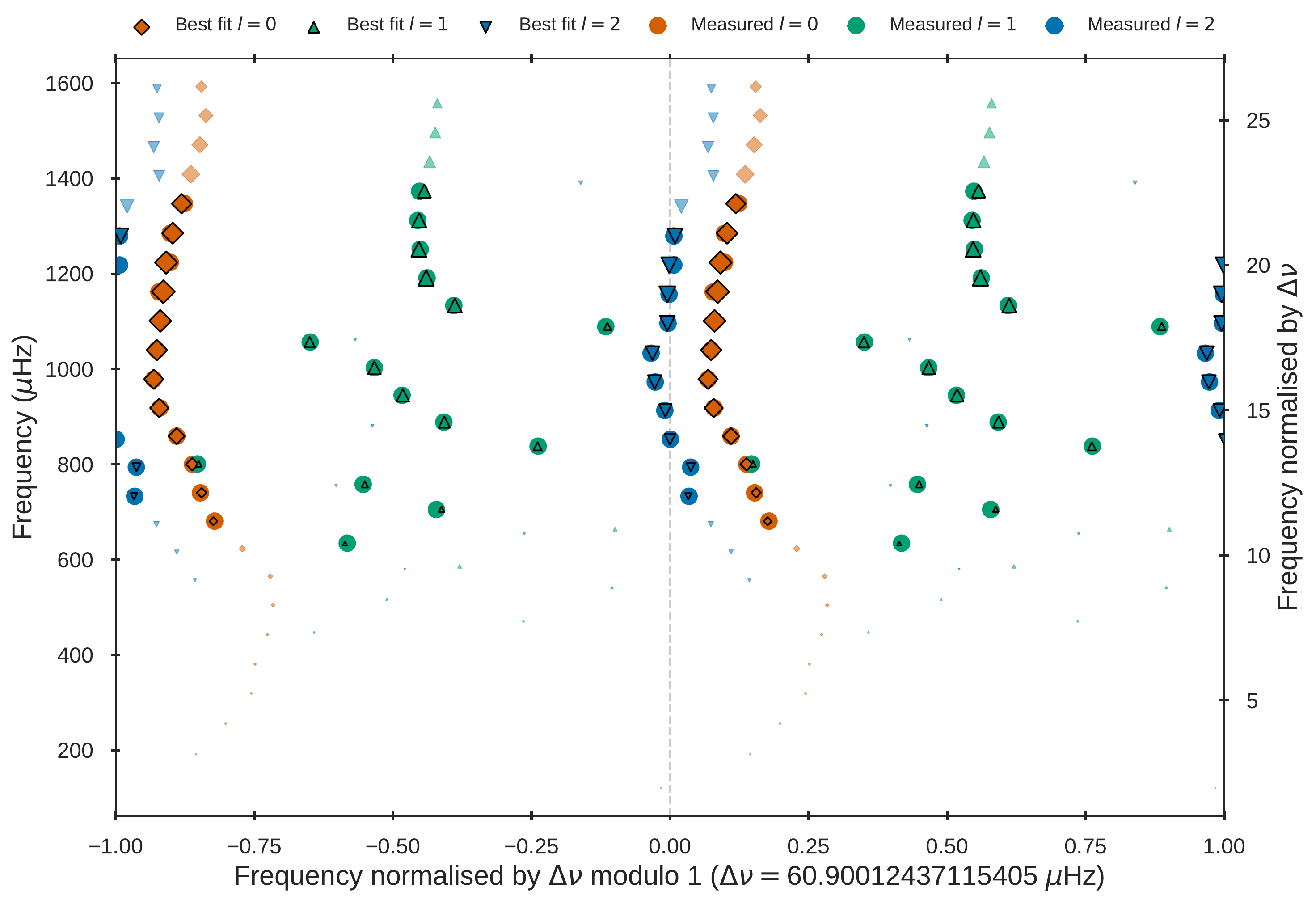}
\caption{\'Echelle diagram of a subgiant validation star of $\dnu\simeq60\mu$Hz.}
\label{fig:SGval60}
\end{figure}
\begin{figure}
\centering
\includegraphics[width=\linewidth]{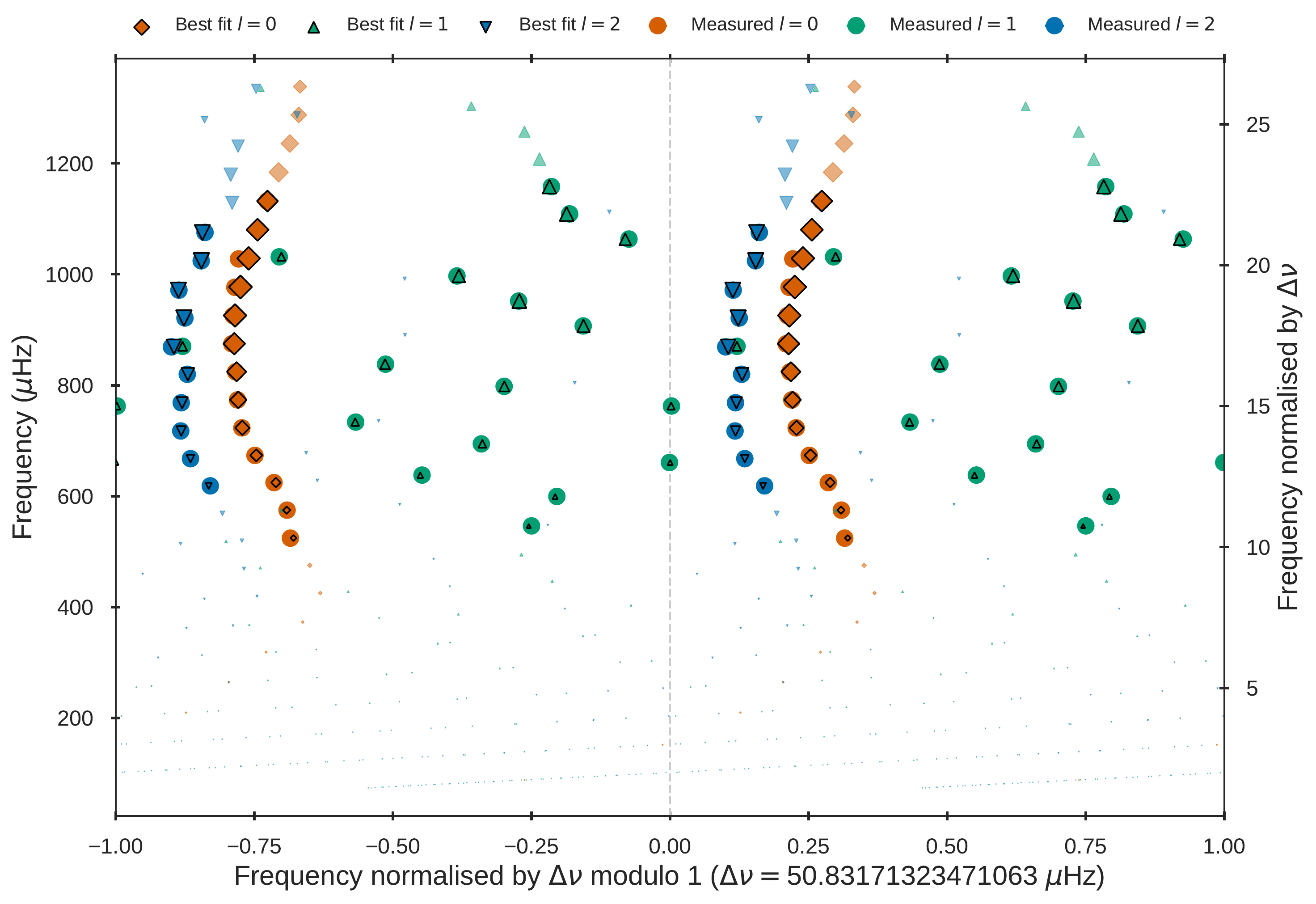}
\caption{\'Echelle diagram of a subgiant validation star of $\dnu\simeq50\mu$Hz.}
\label{fig:SGval50}
\end{figure}
\begin{figure}
\centering
\includegraphics[width=\linewidth]{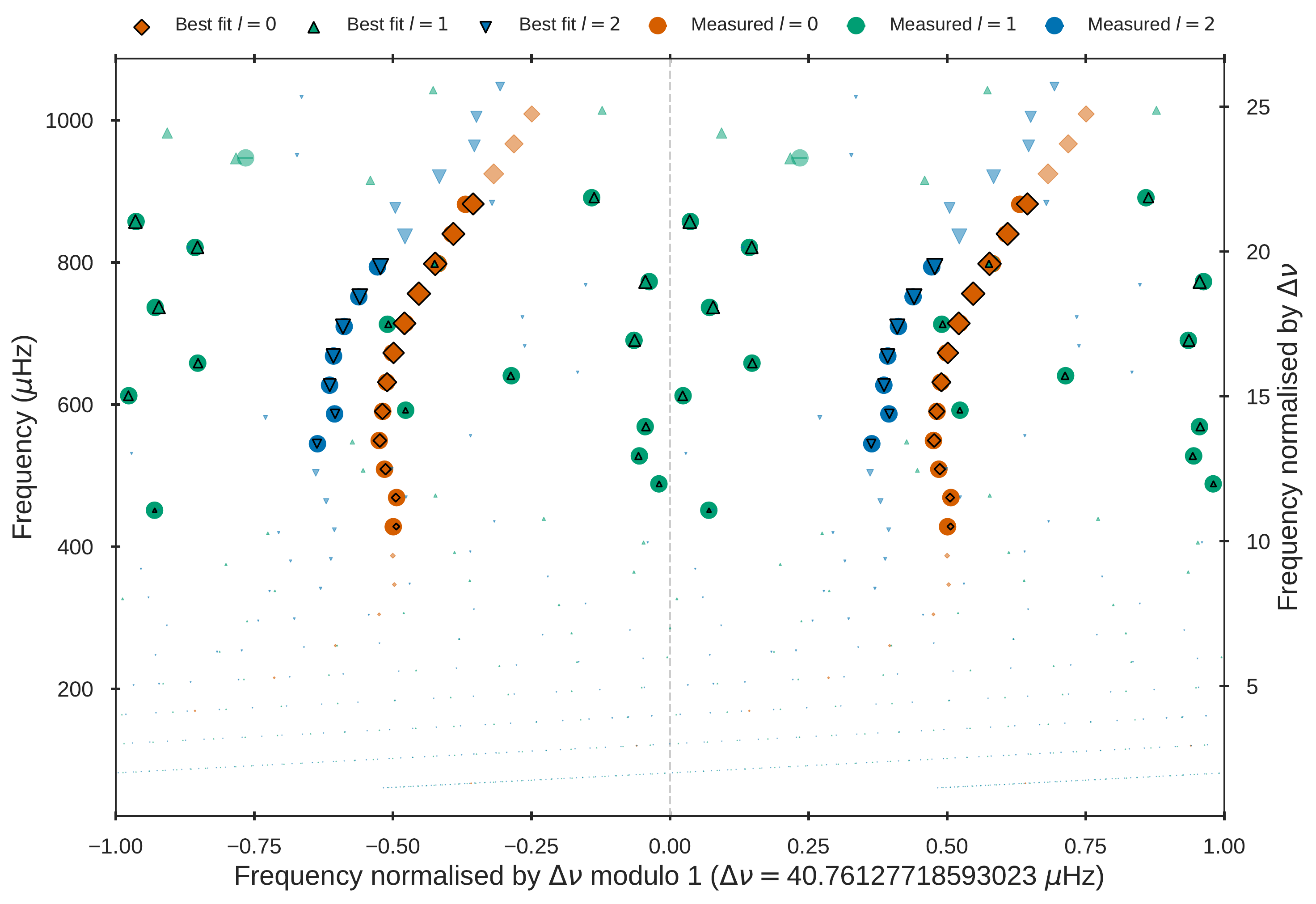}
\caption{\'Echelle diagram of a subgiant validation star of $\dnu\simeq40\mu$Hz.}
\label{fig:SGval40}
\end{figure}
\begin{figure}
\centering
\includegraphics[width=\linewidth]{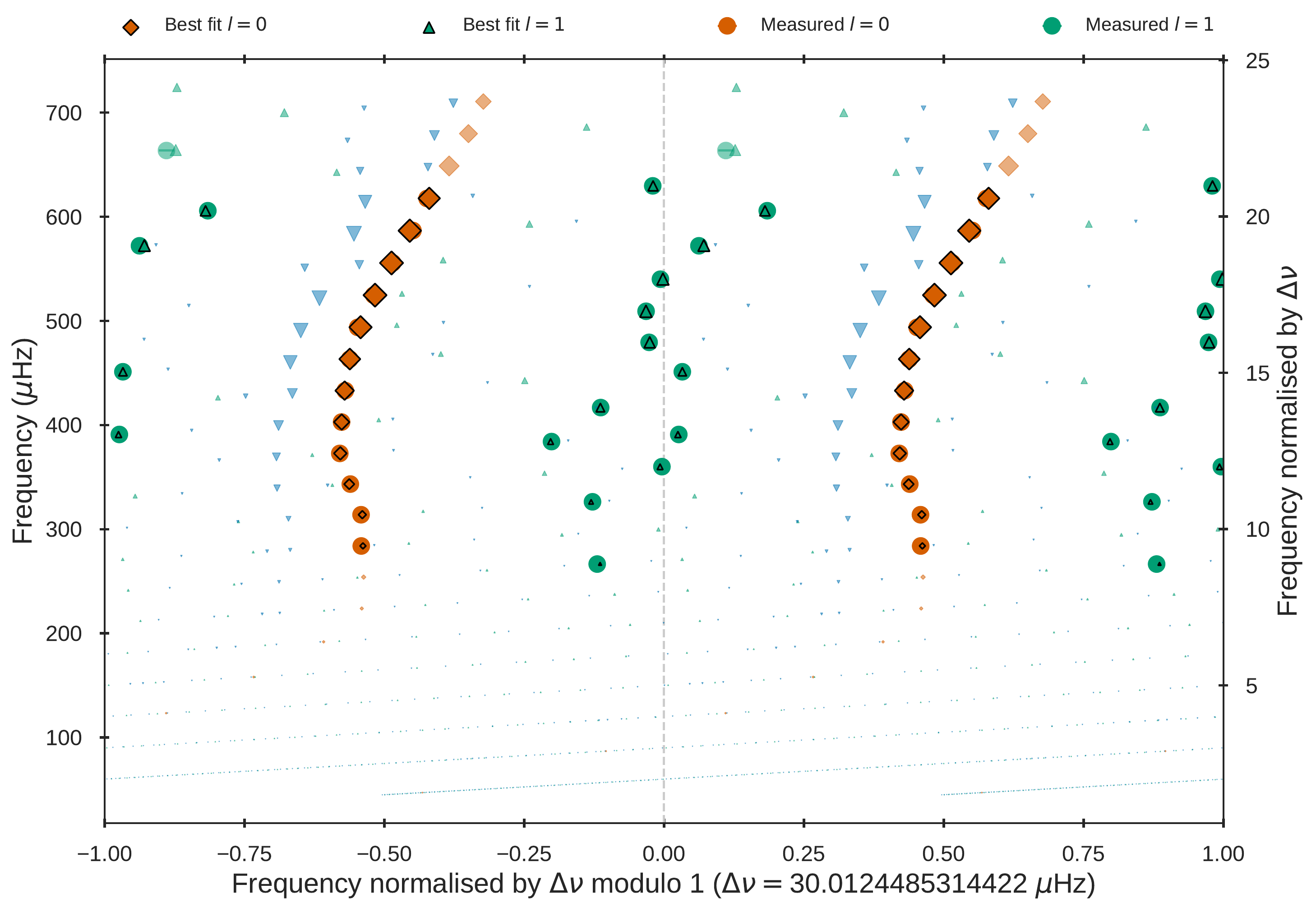}
\caption{\'Echelle diagram of a subgiant validation star of $\dnu\simeq30\mu$Hz.}
\label{fig:SGval30}
\end{figure}
%
% Don't change these lines
\bsp	% typesetting comment
\label{lastpage}
\end{document}